\documentclass[twocolumn,tighten,trackchanges]{aastex631}
\usepackage{graphicx}		
\usepackage{url}
\usepackage[percent]{overpic}
\usepackage{epstopdf}
\usepackage{booktabs}
\usepackage{color}
\usepackage{multirow}
\usepackage{yhmath}
\usepackage{ragged2e}
\usepackage{url}
\usepackage[utf8]{inputenc}
\usepackage{amsmath}
\usepackage{empheq}
\usepackage{glossaries}
\usepackage{tikz}
\usetikzlibrary{calc}

\def\lsim{\mathrel{\raise.3ex\hbox{$<$\kern-.75em\lower1ex\hbox{$\sim$}}}}
\def\gsim{\mathrel{\raise.3ex\hbox{$>$\kern-.75em\lower1ex\hbox{$\sim$}}}}
\def\gtwid{\mathrel{\raise.3ex\hbox{$>$\kern-.75em\lower1ex\hbox{$\sim$}}}}
\def\proptwid{\mathrel{\raise.3ex\hbox{$\propto$\kern-.75em\lower1ex\hbox{$\sim$}}}}
\DeclareUnicodeCharacter{2212}{-}

\def\apj{ApJ}
\def\apjl{ApJL}
\def\apjs{ApJS}
\def\aap{A\&A}
\def\mnras{MNRAS}
\def\pasp{PASP}
\newcommand{\p}{\partial}

\newcommand{\suml}{\sum\limits}

\makeglossaries

\begin{document}

\title{Signatures of Black Hole Spin and Plasma Acceleration in Jet Polarimetry II: Off-Axis Jets}
\shorttitle{Off-Axis Jet Polarization}
\author{Z. Gelles}
\affil{Department of Physics, Princeton University, Princeton, NJ 08540, USA}
\author{A. Chael}
\affil{Princeton Gravity Initiative, Princeton University, Princeton, NJ 08540, USA}
\author{E. Quataert}
\affil{Department of Astrophysical Sciences, Princeton University, Peyton Hall, Princeton, NJ 08540, USA}

\begin{abstract}
We analyze the polarization of optically thin, stationary, axisymmetric black hole jets at scales of order the light cylinder radius. Our work generalizes the face-on results of \cite{Gelles_2025} to arbitrary viewing inclination. 
Due to a combination of geometry and relativistic aberration, the polarization of the jet is not left-right symmetric, and the degree of asymmetry can shed light on both the viewing angle and the plasma bulk Lorentz factor. We show that there is always a radius in the jet at which the polarization transitions from azimuthal to radial; this radius is different along the spine and limb of the jet. We propose metrics that can be used to constrain the black hole spin, inclination angle, and plasma Lorentz factor from these polarimetric signatures, and we discuss the impact of limb-brightening on these measurements. We anticipate that these polarimetric signatures can be studied with current or forthcoming data in M87, NGC 315, NGC 4261, Centaurus A, Cygnus A, and other systems.  Observations of the polarization of the base of the counter-jet in higher inclination sources would provide a particularly promising probe of black hole spin.
\end{abstract}


\section{Introduction}
Radio interferometry serves as a powerful tool for producing spatially resolved images of black hole jets, which are powered by magnetic fields and therefore emit strong synchrotron radiation \citep{blandford_electromagnetic_1977,blandford_relativistic_1979,begelman1984theory}. Radio images have been obtained for nearby systems like Messier 87 (M87; \citealp{curtis1918descriptions,junor1999formation,ly_high-frequency_2007,hada_high-sensitivity_2016,walker_structure_2018,kravchenko2020linear,kino_implications_2022}), as well as more distant radio galaxies and blazars \citep{Pearson_1981,kellermann2004sub,Lister_2005,Marscher_2008_core,hovatta_mojave_2012}, and even cosmological sources \citep{frey2010high}.

Very Long Baseline Interferometry (VLBI) enables observations to directly probe the conditions under which jets are launched, thus connecting the pc-Mpc scale outflows to horizon-scale dynamics. Array upgrades like the next generation Event Horizon Telescope (ngEHT; \citealp{ngEHT1}) and Black Hole Explorer (BHEX; \citealp{johnson_black_2024}) will respectively provide the dynamic range and high angular resolution to image jet launching with exquisite detail. 

From these VLBI images, we hope to infer properties of both the jet and the black hole from which it emerges. Plasma properties within the jet are already well constrained in many sources. For M87 in particular, we know much about how the plasma in the jet accelerates and collimates; this information comes from the observed shape of the jet and from the relative brightness of the counter-jet and forward jet \citep{biretta_hubble_1999,asada_structure_2012,park2019kinematics,kino_implications_2022}. 

It turns out that emission on scales of $\sim 10^{2-4} GM/c^2$ can also still carry information about the spin of the central engine. This is because the jet is believed to be powered by the Blandford-Znajek mechanism \citep{blandford_electromagnetic_1977}, in which the rotational energy of the black hole is directly converted into an electromagnetic outflow. Thus, the electromagnetic fields in the jet (both their magnitude and, in particular, their orientation) depend sensitively on black hole spin. 

Measuring the spins of supermassive black holes is an important yet challenging frontier in modern astrophysics.   Black hole spin connects to a wide range of science questions from the origin of jets, to the hierarchical assembly of massive black holes (and their spin) by accretion and mergers,  to the effects of jet feedback on large-scale structure.    Despite the plethora of AGN whose masses have been well-constrained  \citep{gebhardt2000black,peterson2004black,ferrarese2005supermassive,Vestergaard2006,greene2010precise,Shen_2011,bentz_2015}, the question of spin has remained far more elusive.

For X-ray binaries and luminous quasars, techniques like X-ray reflection spectroscopy \citep{fabian1989x,george1991x,brenneman2006constraining,reynolds2008broad}
and the X-ray continuum method \citep{zhang1997black,shafee2005estimating,mcclintock2006spin} have 
been used to constrain black hole spin in sources accreting at a significant fraction of the Eddington rate.
However, these techniques cannot be used for low-luminosity AGN, which comprise a substantial portion of the galaxies in our local universe, including those with the best studied jets \citep{ho_nuclear_2008}.


Going forward, it will be essential to have an additional model-agnostic method of measuring black hole spin --- and ideally one that does not require spatial resolution at the level of the black hole horizon. Such a requirement then motivates us to look for signatures of spin in the larger-scale structure of black hole jets.

\cite{Gelles_2025} proposed a polarimetric signature of black hole spin for jets observed face-on. In particular, \cite{Gelles_2025} showed that as the magnetic field in the jet is wound up from the rotation of the central black hole, an observer looking down the pole of the jet will see the polarization flip at the light cylinder --- a spin-dependent surface that is typically located $\sim 10-100 GM/c^2$ away from the black hole. Since the polarization is modified by special relativistic effects, this polarization swing depends on the plasma Lorentz factor as well. Jet polarimetry thus offers a promising candidate for measuring black hole spin in conjunction with dynamical properties of the accelerating plasma.


Our goal in this work is to extend the analysis of \cite{Gelles_2025} to arbitrary viewing inclination and understand what properties of jets and the central engine powering them can be measured using jet polarimetry.
In particular, we describe the spatially resolved polarization structure of axisymmetric time-steady black hole jets as a function of black hole spin, observer inclination $i$ and terminal Lorentz factor $\gamma_\infty$. In doing so, we find that polarized jet images can typically be sorted into one of four regimes:
\begin{enumerate}
    \item Face-On Regime (Counter-Jet Present)
    \item Face-On Regime (Counter-Jet Absent)
    \item Low-Inclination Regime $(\gamma_\infty\sin i<1)$
    \item High-Inclination Regime $(\gamma_\infty\sin i>1)$
\end{enumerate}
The face-on regime --- described in detail in \cite{Gelles_2025} and reviewed in the first part of this paper for clarity --- corresponds to observers who view the emission from within the cone of the jet. The low and high inclination regimes --- which are the main focus of this paper and do not depend on the presence of a counter-jet --- correspond to observers who view the emission from outside the cone of the jet. 

In each of these regimes, a polarization swing emerges at a radius that varies inversely with the black hole spin. For the off-axis observers, distinct swings emerge in the spine and in the limbs --- two sub-components of the jet which separate the central emission from the edge emission. However, only one of the two limbs retains a polarization swing, depending on whether one falls into the high-inclination regime or low-inclination regime.   Regardless of which regime a particular jet falls into, strong polarization swings as a function of both distance from the black hole and angle within the jet will always be present. A spatially resolved observation of this polarization swing constrains a combination of black hole spin, observing inclination $i$ and jet Lorentz factor $\gamma_\infty$.
The goal of this paper is to analyze these polarization signatures and discuss some prospects of how they can be studied in practice.

The rest of this paper is organized as follows. In \S\ref{sec:model}, we review the analytic framework of cold, axisymmetric GRMHD that we use to analyze the polarization of black hole jets, and we discuss our ray-tracing procedure. In \S\ref{sec:polfaceon}, we discuss the polarization of black hole jets when viewed from a face-on inclination angle, supplementing the results of \cite{Gelles_2025} to draw an important distinction between observations with a counter-jet and observations lacking a counter-jet. In \S\ref{sec:offaxis}, we then move to the off-axis case.
And in \S\ref{sec:obsforecast}, we discuss the observational implications of this work and offer concrete suggestions for measuring jet properties and black hole spin from jet polarimetry. In Appendix~\ref{app:glossary}, we include a glossary of the relevant terms discussed throughout the paper.

\section{Model}
\label{sec:model}
Black hole jets are visible across the electromagnetic spectrum, with radio observations reflecting the strong magnetic fields used to power the outflows \citep{blandford_relativistic_1979,begelman1984theory,Hada_2019}.  In this section, we describe how to model the dynamics of the plasma in these jets and how to capture their polarized emission patterns. 

Our modeling is motivated by observations of jets like that of M87*, which has been extensively imaged with radio interferometry at a range of scales and frequencies spanning from the optically thick regime ($\lsim 43$ GHz; \citealp{macdonald1968observations,junor1999formation,ly_high-frequency_2007,asada_structure_2012}) to the optically thin regime ($\gtrsim 86$ GHz; \citealp{hada_innermost_2013,hada_high-sensitivity_2016}). In particular, our model is applicable to the ``acceleration zone," where the plasma is speeding up and fieldlines are collimated \citep[e.g.][]{Sanders_1983,Kovalev_2020_transition}. In M87, the recollimation shock --- past which the jet loses its confining pressure and the plasma decelerates --- occurs somewhere near the Bondi radius $\sim 10^5r_g\sim 250\,{\rm pc}$ \citep{Stawarz_2006,asada_structure_2012}.

Outside of M87*, the brightest jets visible on the sky are typically blazars, which are pointed nearly directly at the observer and thus maximize the effects of Doppler boosting (see \citealp{hovatta2020relativisticjetsblazars} for a review). As we will see, our model can be applied to blazars as well, for which many observations directly resolve parts of the acceleration zone \citep{Marscher_2008_core}. In all cases, we will restrict our theoretical analysis here to the optically thin limit.

\subsection{Jet Structure}
In order to compute the polarized emissivity, we need to know three things about the jet: its shape, its magnetic field profile, and the velocity of the plasma therein. In principle, all three of these quantities can be self-consistently obtained by solving the equations of General Relativistic Magnetohydrodynamics (GRMHD). For our purposes, we will turn to approximate GRMHD solutions, which broadly reproduce  observed properties of jets.

In particular, we will adopt the framework of \cite{Gelles_2025}, which adapts known solutions of cold, axisymmetric GRMHD \citep[as originally developed by, e.g., ][]{weber_angular_1967,michel_rotating_1973,mestel1979axisymmetric,blandford_hydromagnetic_1982,phinney_theory_1984,camenzind_hydromagnetic_1986,camenzind_hydromagnetic_1987} to our setting of interest. We review this framework briefly below, referring the reader to \cite{Gelles_2025} for a much more detailed exposition.

This framework describes a jet anchored to a Kerr black hole of mass $M$ and dimensionless spin $0\leq a\leq 1$. The shape of the jet is captured by a stream function $\psi$, which labels individual magnetic fieldlines and is written in terms of a collimation parameter\footnote{Note that we have rescaled the stream function from \cite{Gelles_2025} by a factor of $r_+^{-p}$. This ensures that the last fieldline to thread the horizon is labeled by $\psi=1$.} $0\leq p<2$:\begin{align}
    \psi&=\left(\frac{r}{r_+}\right)^p(1-\cos\theta),
\end{align}
where $r_+/M=1+\sqrt{1-a^2}$ is the radius of the outer horizon. The parameter $p$ sets the steepness of the fieldlines via the parametric correspondence\footnote{Here, $(r,\theta,\phi)$ are spherical coordinates and $(R,Z,\phi)$ are cylindrical coordinates. See Appendix~\ref{app:coordinates} for more details.}\begin{align}
    Z\propto R^{\frac{2}{2-p}}.
\end{align}

Unless otherwise specified, we will take $p=0.75$, thus producing a collimated jet with $Z\propto R^{1.6}$. This choice of $p$ is well motivated by numerical simulations of jets \citep{mckinney_disk-jet_2007,tchekhovskoy_simulations_2008,nakamura_parabolic_2018}, as well as radio observations of systems within the acceleration zone like M87*, \citep{asada_structure_2012,hada_innermost_2013,nakamura_parabolic_2018,park_jet_2021}, NGC 315 \citep{park_discovery_2024}, and many other known blazars \citep{pushkarev2017mojave,burd2022dual}. Furthermore, the results presented in this paper will correspond to emission along a single magnetic fieldline. Unless otherwise specified, the fieldline of interest will be that which intersects the black hole's event horizon at the midplane, thus giving $\psi=1$.

Regardless of $p$, the constancy of $\psi$ along this fieldline ensures that the poloidal components of the magnetic field decay as
\begin{align}
\label{eq:fieldexpansion}
    B^Z\propto R^{-2},\quad B^R\propto Z^{-1}R^{-1}.
\end{align}
$B^R$ thus decays faster than $B^Z$ in a collimated jet.

The toroidal component of the magnetic field, as well as the fluid four-velocity, can then be computed explicitly from a quartic ``wind equation," which encapsulates the conservation of energy-momentum in GRMHD. We follow \cite{Gelles_2025} and expand around the strongly magnetized (force-free) limit, where the solutions to the wind equation become analytic. In this limit, the toroidal magnetic field decays exactly as $R^{-1}$, and the fluid velocity is given by pure $\vec{E}\times\vec{B}$ drift, which accelerates the plasma to arbitrarily large Lorentz factors. 

The proportionality constant between the poloidal and toroidal components of $B$ is expressed in terms of the ``fieldline rotation rate" $\Omega_F$, which describes how fast the magnetic fieldlines are rigidly rotating around the black hole. One can show from regularity at the horizon that $\Omega_F$ must be linear in the black hole spin $a$, thus endowing the magnetic field with spin dependence \citep{znajek_black_1977}. In particular, the magnetic field in the strongly magnetized regime satisfies\footnote{In Eq.~\ref{eq:lcdefn}, the hat on $B^{\hat{\phi}}$ indicates that the components are expressed in an orthonormal cylindrical basis, so that $B^{\hat{\phi}}\equiv \sqrt{B^\phi B_\phi}$. See Appendix~\ref{app:coordinates} for details.}\begin{align}
\label{eq:lcdefn}
\frac{B^{\hat{\phi}}}{B^Z}\approx R\Omega_F\equiv \frac{R}{R_{\rm LC}}.
\end{align}
Here, $R_{\rm LC}\equiv \Omega_F^{-1}$ is the light cylinder radius in flat space--- the point at which the magnetic fieldlines rotate at the speed of light. Interior to the light cylinder, the magnetic field is predominantly poloidal. Outside of the light cylinder, the magnetic field winds up and is predominantly toroidal. In GRMHD, $\Omega_F\propto a$ (with dimensionless prefactor fixed by regularity at the horizon), so the light cylinder radius scales as $R_{\rm LC}\propto a^{-1}$ and therefore directly traces the black hole spin. 

To capture the leading order deviation from the force-free regime, \cite{Gelles_2025} modifies the velocity profile of the jet by smoothly capping the Lorentz factor of the plasma at a terminal value $\gamma_\infty$, which is reached asymptotically far from the black hole. Physically, $\gamma_\infty$ is set by the ratio of magnetic energy to the plasma inertial energy, as can be shown in both analytic theory \citep{michel_relativistic_1969,goldreich_stellar_1970,camenzind1989hydromagnetic,li_electromagnetically_1992,beskin_mhd_1998,pu_properties_2020} and numerical simulations \citep{Komissarov_2007,Ripperda_2022}.

In sum, our model describes the dynamics of AGN jets using approximate solutions to cold, axisymmetric GRMHD. To generate these solutions, one needs to specify only three input parameters: the collimation parameter $p$, the terminal Lorentz factor $\gamma_\infty$, and the black hole spin $a$. In many observed jets, total-intensity observations have allowed for constraints to be placed on the first two parameters \citep{biretta_hubble_1999,hovatta_doppler_2009,Homan_2015,Kutkin_2019,pushkarev2017mojave,jorstad2017kinematics,kino_implications_2022,park_jet_2021}. Specifically, the shape of the jet can simply be inferred from the projected image, while the terminal Lorentz factor can be measured from a combination of apparent superluminal motion \citep{rees1966appearance} and jet-counterjet brightness ratios \citep{bridle1984extragalactic}. We will spend the remainder of this paper explaining how jet polarization can be used to constrain black hole spin, jet Lorentz factor, and the observer inclination angle.   

\subsection{Ray-Tracing}
\label{sec:raytracing}
To generate observable quantities from our model of the jet dynamics, we need to perform ray-tracing that connects the observer to the source, thus projecting the three-dimensional jet onto a two-dimensional image. We describe our procedure for doing so below.

We assume that the observer is located infinitely far away from the black hole at an inclination angle $i$ relative to the black hole's spin axis (which is aligned with that of the jet). The observer's screen, which is oriented perpendicular to the line of sight, can be described by Bardeen coordinates $\{\alpha,\beta\}$. These coordinates are aligned with the observer's horizontal and vertical directions respectively. The screen can also be described in polar coordinates $\{b,\varphi\}$, where\footnote{Note that $\varphi=0$ aligns with the $+\hat{\alpha}$ direction on the image plane, i.e. west.}\begin{align}
    b=\sqrt{\alpha^2+\beta^2},\qquad \varphi=\arg(\alpha+i\beta).
\end{align}
$b$ is typically referred to as the ``impact parameter." The projection of the black hole spin axis points up (i.e. in the $+\hat{\beta}$ direction) on the observer's screen.

The bulk spacetime can then be described in terms of Cartesian coordinates $\{X,Y,Z\}$ or cylindrical coordinates $\{R,\phi,Z\}$ (among others). We will use both of these coordinate systems throughout this work, all of which are defined in the usual way but are reviewed for clarity in Appendix~\ref{app:coordinates}. We assume that the observer is placed along the negative $Y$ axis (i.e. at $\phi=-\pi/2$), so that $\varphi=0$ and $\phi=0$ are aligned.

In this paper, we will work in flat space (unless stated otherwise), where ray-tracing is simple since null geodesics are straight lines. Such an approximation is justified since we are concerned with jet emission far from the black hole, where general relativistic corrections of order $\mathcal{O}(M/r)$ are very small. Therefore, a ray originating from the point $(X_0,Y_0,Z_0)$ will hit the observer's screen at\begin{align}
\label{eq:raytracingformula}
    \alpha=X_0,\qquad \beta=Z_0\sin i+Y_0\cos i.
\end{align}
Along a collimated jet, one has $Z_0\gg Y_0$, so the observed ``length" of the jet is given by $\beta\approx Z_0\sin i$. In the face-on regime, $\sin i\ll 1$, and the projected size of the jet instead becomes the simple relation $b=R$. For this reason, we will describe our results in the face-on regime in terms of $b$, and we will describe our results in the off-axis regime in terms of $\beta$.

For our purposes, however, we need the inverse of Eq.~\ref{eq:raytracingformula}; we must be able to compute which part of the jet is projected onto a given point on the observer's screen. This inversion is not generically analytic (unless the jet is purely monopolar or parabolic) but can be reduced to a simple one-dimensional root find. To see this, note that the wave-vector $\vec{k}$ has Cartesian components\begin{align}
    k_X=0,\quad k_Y=-\sin i,\quad k_Z=\cos i,
\end{align}
meaning that geodesics in flat space are parameterized by \begin{align}
    (X,Y,Z)&=(X_0,Y_0-t\sin i, Z_0+t\cos i)
\end{align}
with $t\in(-\infty,\infty)$. For known screen coordinates $\{\alpha_0,\beta_0\}$, the emitted ray will thus intersect the fieldline $\psi_0$ when\begin{align}
    \psi_0=\psi\left(\alpha_0,\frac{\beta_0}{\cos i}-t\sin i,t\cos i\right).
\end{align}
Given a stream function $\psi(x,y,z)$, the above expression can be solved for $t$ using standard numerical methods, thus providing an explicit map between points in the jet and points on the observer's image.\footnote{In \cite{Gelles_2025}, the software \texttt{kgeo} \citep{chael_kgeo_2023} is used to perform fully general relativistic ray-tracing. That level of accuracy is not necessary for the analysis in this paper.}

\begin{figure*}[t]
    \centering
    \includegraphics[width=\textwidth]{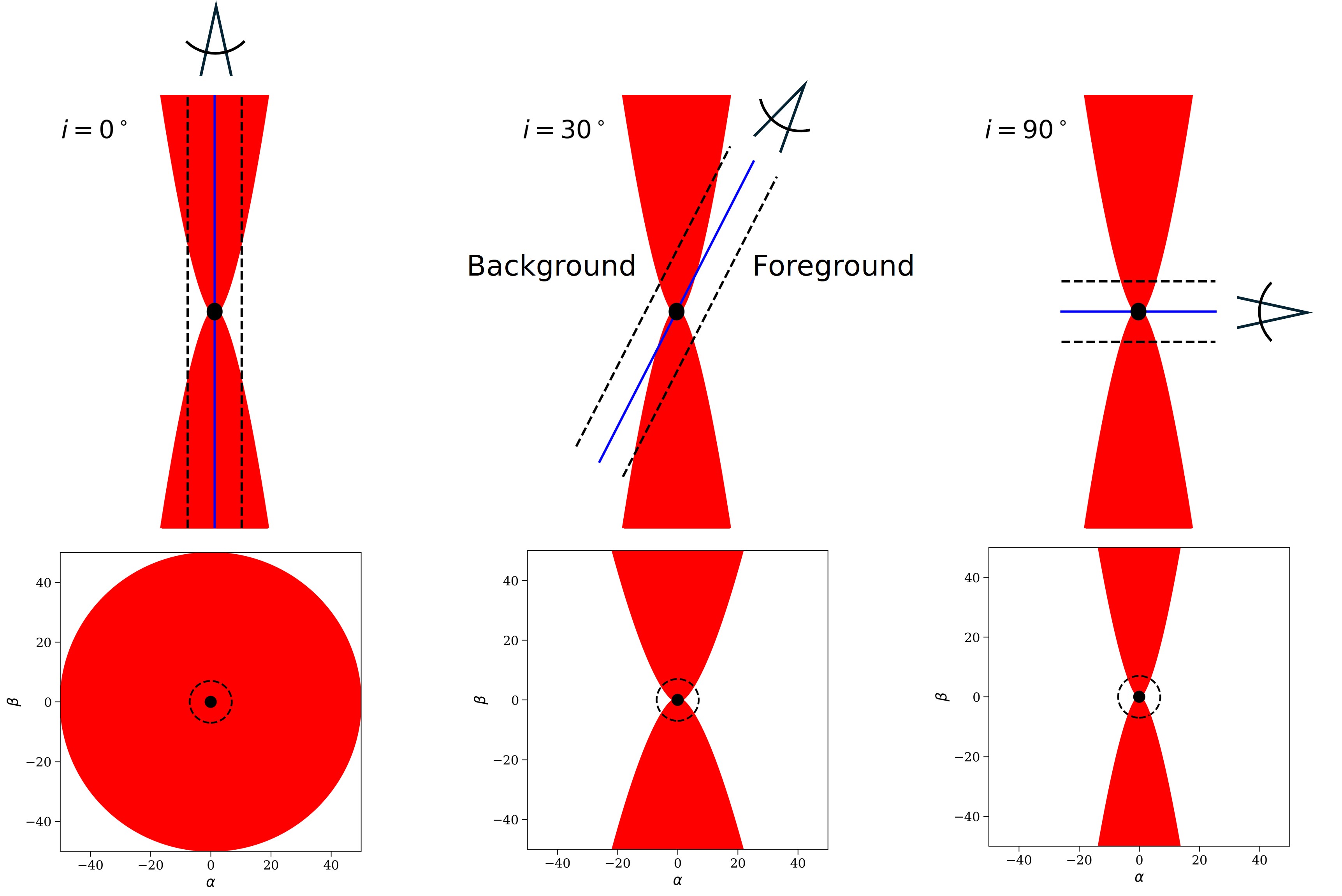}
    \caption{Three different viewing geometries $(i=0^\circ, i=30^\circ, \,i=90^\circ)$ for a sample jet, with coordinate axes plotted in units of $M$. For each inclination angle, the position of the observer is shown on top, with the corresponding projected image shown below. The filled in black circle corresponds to the black hole, and the black dashed line demarcates an impact parameter of $b=7M$. The solid blue line demarcates the origin of the observer's screen. In practice, the $i=90^\circ$ case is typically not observable due to lack of Doppler boosting, but it is conceptually useful.}
    \label{fig:jetangles}
\end{figure*}

The picture of ray-tracing looks qualitatively very different for observers in the face-on regime and observers in the off-axis regime. For emission originating from a polar angle of $\theta_{\rm emit}$, observers with $\sin i < \sin \theta_{\rm emit}$ are inside the effective cone of the jet. But when $\sin i$ exceeds $\theta_{\rm emit}$, the observer is pushed outside of the jet cone and begins to see the jet collimate, with higher inclinations seeing a progressively narrower projection of the full jet profile. Indeed, the critical emission angle $\sin\theta_{\rm emit}=\sin i$ corresponds to the largest contour of constant $Z$ enclosing the origin, as can be seen by evaluating Eq.~\ref{eq:raytracingformula} with $\alpha=\beta=0$. In conical jets (e.g. \citealp{blandford_relativistic_1979}), $\theta_{\rm emit}$ is constant and identified with the jet's ``opening angle."

Furthermore, rays connecting observer to source generically intersect the emission region twice, as can be seen in Figure~\ref{fig:jetangles}. For observers in the face-on regime (left panel of Figure~\ref{fig:jetangles}), each ray inherits one intersection with the forward jet and one intersection with the counter-jet. But for off-axis observers (middle and right panels of Figure~\ref{fig:jetangles}), each ray inherits two intersections in the same hemisphere. 

For these reasons, it is important for us to analyze the polarization of face-on images (viewed from within the jet cone) and off-axis images (viewed from outside the jet cone) using  different frameworks. It is often possible to infer the inclination of a jet using the same techniques as those used to measure $\gamma_\infty$ \citep{urry_1995}. The inclination of the M87* jet, for example, has been well constrained to satisfy $i\approx 17^\circ$ \citep{mertens_kinematics_2016}. 

\subsection{Polarization}
\label{sec:polarizationsub}
Once we know the shape of the jet and how it projects onto the observer's two-dimensional screen, we then need to compute the observed polarization as a function of the screen coordinates. 

To do so, we use the fact that the polarization vector $\vec{f}$ is always perpendicular to the emitted wave-vector and the local magnetic field in the fluid frame \citep{Rybicki_Lightman}. In other words, \begin{align}
\label{eq:kcrossBfluid}
    \vec{f}|_{\rm fluid}\propto (\vec{k}\times\vec{B})|_{\rm fluid}.
\end{align}
Since $\vec{f}$ is a vector, it transforms non-trivially under Lorentz boosts and will therefore differ between the fluid frame and the lab frame. In Appendix~\ref{app:polidentities}, we evaluate the above expression to reveal that in GRMHD, the polarization in both of these frames can be captured concisely via the formula
\begin{align}
\label{eq:polcross}
   \vec{f}\propto \vec{k}\times\vec{B}+\left(\frac{v^\perp}{v_{\rm FF}^\perp}\right)\vec{E},
\end{align}
where $v$ is the magnitude of the plasma MHD velocity, $v_{\rm FF}$ is the magnitude of the force-free drift velocity, and $\vec{E}$ is the electric field. Specifically, the electric field in the fluid frame is zero and in the lab frame is equal to \begin{align}
    \vec{E}|_{\rm lab}&=R\Omega_F\hat{\phi}\times\vec{B}.
\end{align}
Note that Eq.~\ref{eq:polcross} is exact (up to the proportionality constant) and captures the Lorentz transformation of the polarization from fluid frame to lab frame, thus encoding all special relativistic effects we need.

From here, we then parallel transport $f$ along the trajectories derived in \S\ref{sec:raytracing}. In flat space, this procedure is completely analytic and is most easily expressed in terms of Cartesian coordinates. Following the sign conventions from \cite{himwich_universal_2020} and measuring the Electric Vector Position Angle (EVPA) from the $+\hat{\beta}$ axis, we have\begin{align}
\label{eq:polgen}
    \tan({\rm EVPA})&=-\frac{f^X}{f^Y\cos i+f^Z\sin i}.
\end{align}

If emission along a ray comes from multiple sources, then the net polarization angle is determined via the Stokes parameters $\{Q,U\}$, which are additive in the optically thin limit and are related to the EVPA via:\begin{align}
    Q+iU=Pe^{2i\,{\rm EVPA}},
\end{align}
with $P$ the magnitude of the polarized intensity. To compare $P$ between the foreground and background of our model, we assume that the polarized emissivity is proportional to the plasma density, with the plasma density set by equipartition of magnetic and thermal energy \citep{blandford_relativistic_1979}. The appropriate Doppler factor is also included to enhance the polarized intensity in regions beamed towards the observer (see \citealp{Gelles_2025}).

\section{Face-On Polarization}
\label{sec:polfaceon}
Throughout the rest of this paper, we will expand Eq.~\ref{eq:polgen} in various regimes to understand the spatial evolution of jet polarization for generic viewing geometries and source profiles. As mentioned in \S\ref{sec:raytracing}, the face-on and off-axis observers see qualitatively different images, so it is important to analyze them separately. To this end, we will begin by analyzing the face-on polarization structure before presenting novel off-axis results. 
Here and throughout the rest of this paper, the ``face-on" regime specifically applies when the emission on scales of interest is viewed from within the cone of the jet ($\sin i<\sin \theta_{\rm emit}$). This low-inclination limit is treated in great detail in \cite{Gelles_2025}, the results of which are reviewed and supplemented below.

\subsection{Computation of the EVPA}
When the observer views the the emission from within the cone of the jet, an axisymmetric source will produce an axisymmetric image. As such, we can describe the entire face-on polarization pattern by simply analyzing the EVPA along the contour of $\varphi=\pi/2$ (i.e. 12 o'clock on the image). 

Along this contour, the Cartesian unit vectors align with their cylindrical counterparts: $e_X||-e_{\hat{\phi}}$ and $e_Y||e_R$, so that Eq.~\ref{eq:polgen} becomes\begin{align}
    \label{eq:polfaceon2}
    \tan({\rm EVPA})|_{\rm face-on}&=\frac{f^{\hat{\phi}}}{f^R}.
\end{align}
The above expression cleanly separates into a poloidal denominator with a toroidal numerator. This means that as the magnetic field winds up from poloidal to toroidal, the polarization will correspondingly swing $90^\circ$ from azimuthal (EVPA=$\pi/2$) to radial (EVPA=$0$). To compute the explicit form of this swing, we expand in terms of the electromagnetic fields using Eq.~\ref{eq:polcross}, which gives
\begin{align}
\label{eq:evpafaceon}
    \tan({\rm EVPA})|_{\rm face-on}&\approx \frac{B^R}{-B^{\hat{\phi}}+\frac{v^\perp}{v_{\rm FF}^\perp} E^R},
\end{align}
where we set $i=0$ everywhere (as appropriate for the face-on regime). As one can see, all $v$ dependence enters only through the denominator of Eq.~\ref{eq:aberrationspine}, meaning that a nonzero velocity effectively slows down the polarization wind-up. This is exactly what we found in \cite{Gelles_2025} via explicit Lorentz transformation.

Now, Eq.~\ref{eq:evpafaceon} must be evaluated separately in the forward jet and in the counter-jet, and there is potential for significant cancellation between the two. Indeed, the observed polarization will depend sensitively on the brightness of the counter-jet relative to that of the forward jet. Below, we discuss the observed polarization in cases where a counter-jet is completely visible, and then subsequently in cases where it is not.

\subsection{Counter-Jet Present}
\label{sec:counterjetsec}
When a counter-jet is present, polarized emission will be subject to cancellation, as can be seen by expanding the Stokes parameters:\begin{align}
    Q|_{\rm face-on}&=1-\frac{2(f^{\hat{\phi}})^2}{(f^{\hat{\phi}})^2+(f^{R})^2}
    \\ U|_{\rm face-on}&=-\frac{2f^{\hat{\phi}}f^{R}}{(f^{\hat{\phi}})^2+(f^{R})^2}.
\end{align}
Since $B^R$ flips sign from forward jet to counter-jet, then so too does $f^{\hat{\phi}}$ and hence Stokes $U$. This means that deep inside the light cylinder, where Doppler factors are close to unity, Stokes $U$ can cancel between the forward jet and counter-jet emission, creating a radial polarization pattern in spite of a predominantly radial magnetic field\footnote{In fact, as shown in \cite{Gelles_2025}, general relativistic effects can actually enhance the synchrotron pitch angle in the counter-jet, effectively making the counter-jet even \emph{brighter} than the forward jet. For this reason, the plots in this section --- unlike the rest of the paper --- are plotted using the fully general relativistic polarization model described in \cite{Gelles_2025}.}. 

\begin{figure}[h]
    \centering
    \includegraphics[width=0.45\textwidth]{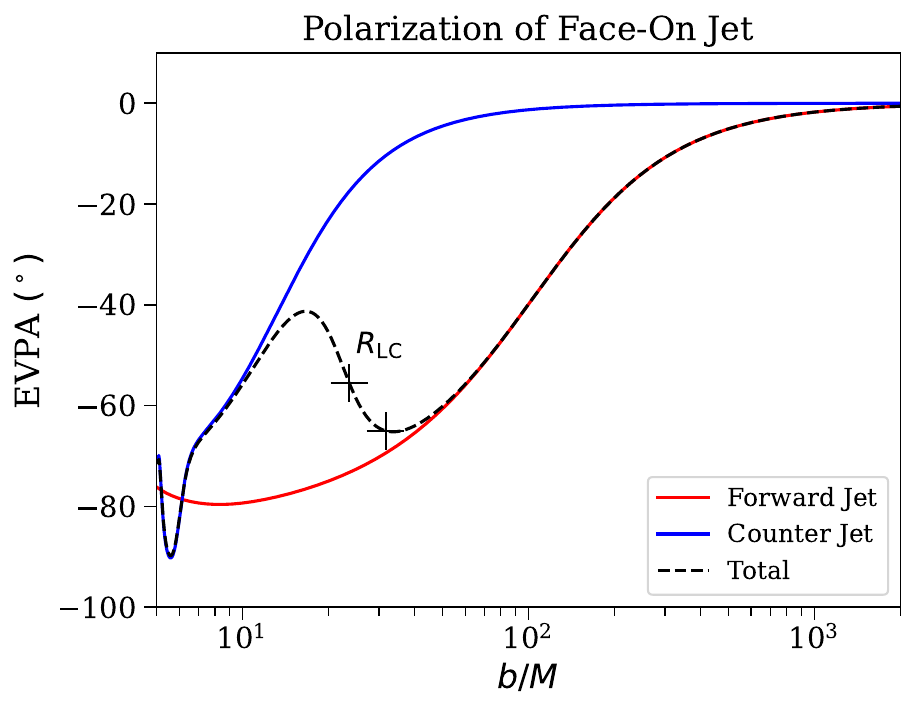}
    \caption{Polarization of a face-on jet with $a=0.5$ and $\gamma_\infty=5$, viewed along the image contour $\varphi=\pi/2$ (i.e. 12 o'clock). The jet and counter-jet each produce different polarization patterns, with the forward jet only becoming the dominant source of polarization outside the light cylinder. When a counter-jet is present, the net polarization therefore swings sharply at the light cylinder. The two crosses mark the lensed positions of the light cylinder in the forward jet and light cylinder in the counter-jet respectively. Note that this calculation includes GR lensing to match Figure 10 of \cite{Gelles_2025}, while the Figures in other sections of this paper use a flat space approximation.}
    \label{fig:faceonCJfig}
\end{figure}

But as the light cylinder is approached, the plasma begins to accelerate, causing the counter-jet emission to become Doppler de-boosted. This means that Stokes $U$ no longer cancels, forcing the polarization to swing back to its azimuthal value before winding back up with the magnetic field in the forward jet. Thus, when a counter-jet is present, we expect face-on images to display a sharp polarimetric swing at the light cylinder, as shown in Figure~\ref{fig:faceonCJfig}. This swing was first predicted in \cite{Gelles_2025} (cf. their Figure 10).

However, it is reasonable to expect that some face-on sources might \emph{not} feature a visible counter-jet. In this case, the polarization pattern will look quite different than that of Fig.~\ref{fig:faceonCJfig}. This scenario was not treated in \cite{Gelles_2025}, so we analyze it in detail below.

\subsection{Counter-Jet Absent}
When a counter-jet is absent, the EVPA of the face-on image will be described by Eq.~\ref{eq:evpafaceon} along the forward jet alone. Without cancellation between the forward jet and counter-jet, the polarization will not swing abruptly at the light cylinder anymore; instead, the polarization will gradually transition from azimuthal to radial over a wide range of impact parameters. 

\begin{figure*}[t]
    \centering
    \includegraphics[width=\textwidth]{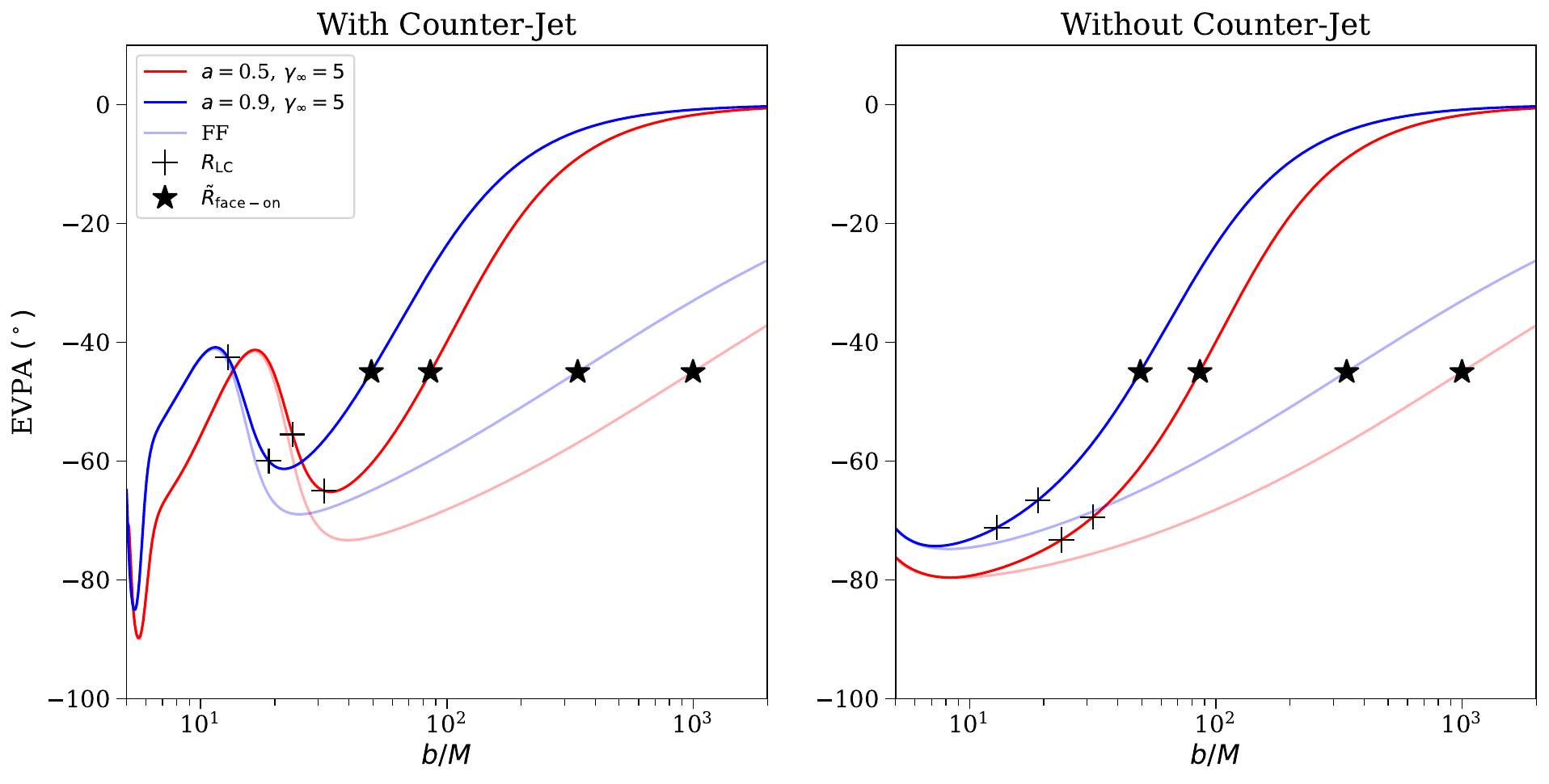}
    \caption{Polarization of a face-on jet, again including fully general relativistic effects. On the left, the results are shown when a counter-jet is present (thus matching Figure~\ref{fig:faceonCJfig}), and on the right, the results are shown when a counter-jet is not present. In each panel, polarization curves are plotted for two different spins ($a=0.5$ and $a=0.9$) and terminal Lorentz factors $(\gamma_\infty=5$ and force-free). When a counter-jet is present, the polarization curve always shows a sharp swing at the light cylinder (marked with black crosses), regardless of terminal Lorentz factor. But when the counter-jet is not present, the polarization swings more modestly at the true light cylinder, meaning that $\tilde{R}$ (shown as a star) can serve as a more easily identifiable metric.}
    \label{fig:cjcomparefig}
\end{figure*}

This distinction is present in Figure \ref{fig:faceonCJfig} and is elaborated on in Figure~\ref{fig:cjcomparefig}. This figure shows that when a counter-jet is absent, the polarization still begins to swing from azimuthal to radial at the light cylinder. But the swing is no longer sharp, rendering its location more ambiguous and harder to identify.

To this end, it is useful for us to define an unambiguous ``effective swing location" along the jet. In particular, we introduce an emission radius $\tilde{R}$ --- a parameter which we will use throughout the rest of this work --- to coincide with this effective swing location:\begin{align}
\label{eq:Rtildedef}
    \tilde{R}\equiv \text{cylindrical radius of pol. swing}.
\end{align}
Emission from $\tilde{R}$ produces a polarization swing on the observer's image at a Bardeen coordinate $\tilde{\beta}$, or equivalently at an impact parameter $\tilde{b}$:\begin{align}
    \tilde{\beta}&\equiv \text{vertical image coordinate of pol. swing},
    \\
    \tilde{b}&\equiv \text{impact parameter of pol. swing}
\end{align}
with $\tilde{R}$ and $\tilde{\beta}/\tilde{b}$ related via the ray-tracing formula in Eq.~\ref{eq:raytracingformula}.

Throughout this work, we will use the parameter $\tilde{R}$ to identify the location of the polarization swing. But the value of the EVPA at which the polarization swings will depend heavily on the observer's inclination, so we will include subscripts on the parameter $\tilde{R}$ to denote its value in the different inclination regimes.

For the case of the face-on viewing geometry discussed in this section, we find it practical to identify $\tilde{R}_{\rm face-on}$ --- the ``swing location" of the face-on jet --- with the point at which the EVPA has wound up exactly halfway from azimuthal (${\rm EVPA}=\pm 90^\circ)$ to radial $({\rm EVPA}=0^\circ)$:\begin{align}
\label{eq:faceonRtilde}
    {\rm EVPA}|_{\tilde{R}_{\rm face-on}}=\pm 45^\circ.
\end{align}

Regardless of whether a counter-jet is present, \emph{all} face-on images of black hole jets show the polarization wind-up with the magnetic field, and hence will possess a unique value of $\tilde{R}_{\rm face-on}$ as defined by Eq.~\ref{eq:faceonRtilde}. The location of $\tilde{R}_{\rm face-on}$ for multiple face-on jets is shown in Figure~\ref{fig:cjcomparefig}, where the impact parameter corresponding to emission from $\tilde{R}_{\rm face-on}$ simply satisfies $\tilde{b}_{\rm face-on}=\tilde{R}_{\rm face-on}$.

\begin{figure*}[t]
    \centering
    \includegraphics[width=\textwidth]{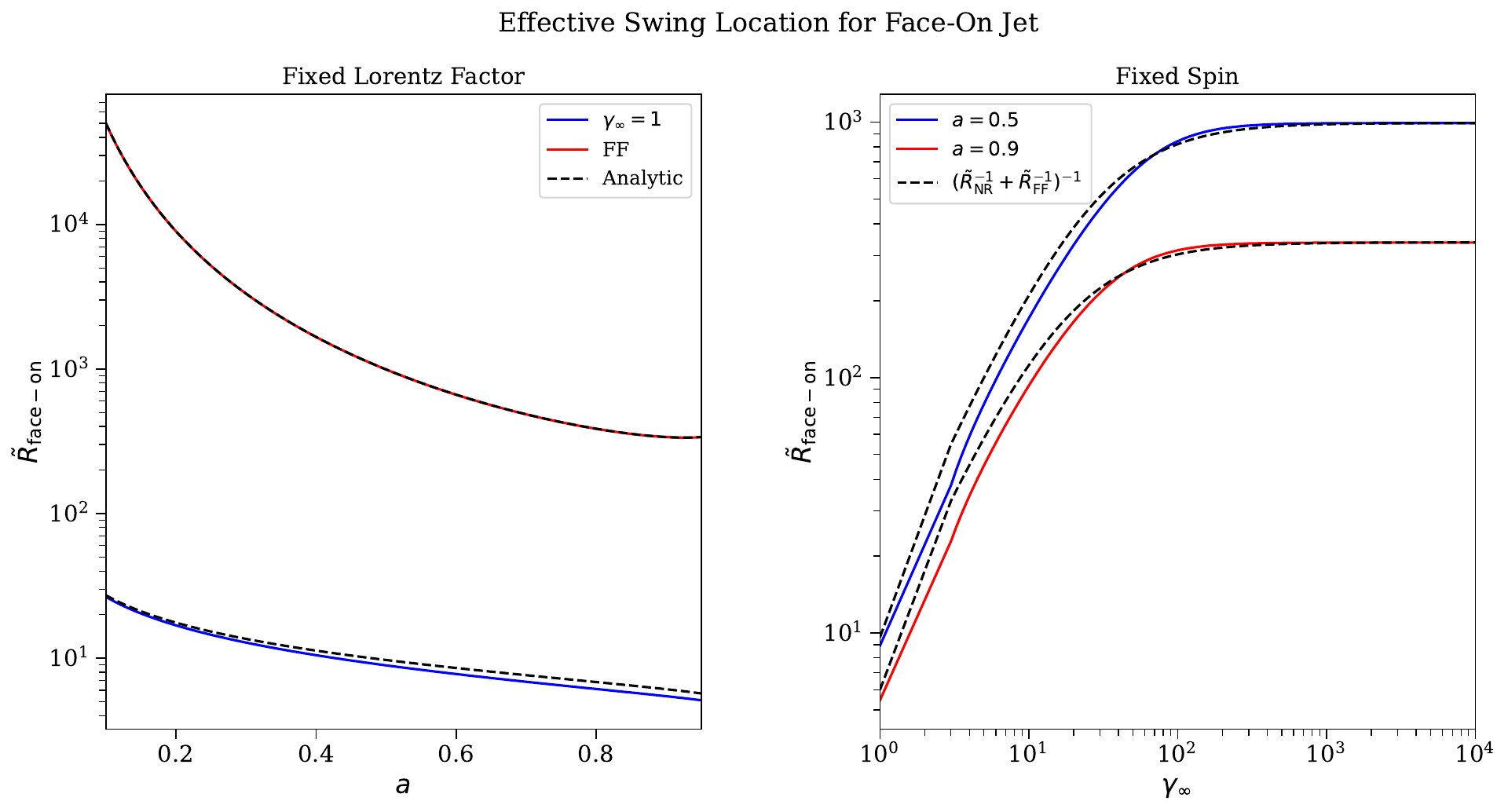}
    \caption{Effective EVPA swing location $(\tilde{R}_{\rm face-on})$ as a function of both spin (left panel) and terminal Lorentz factor (right panel) for a sample jet. Analytic solutions exist in the non-relativistic ($\gamma_\infty\sim 1$) regime and the force-free ($\gamma_\infty\gg 1$) regime, which are shown as blacked dashed lines in the left panel. These two analytic solutions can then be stitched together using Eq.~\ref{eq:seriesfaceon}, which is overplotted as black dashed lines in the right panel. In all cases, $\tilde{R}_{\rm face-on}$ depends strongly on black hole spin and $\gamma_\infty$.}
    \label{fig:faceonscaling}
\end{figure*}

To compute the dependence of $\tilde{R}_{\rm face-on}$ on the structure of the jet, we simply have to invert Eq.~\ref{eq:evpafaceon}. This inversion is detailed in Appendix~\ref{app:faceon} and can be completed analytically in both the non-relativistic ($\gamma_\infty \sim 1$) regime and ultra-relativistic/force-free ($\gamma_\infty\gg 1$) regime. Making our default assumption that $\psi=1$ and the fieldline of interest threads the horizon at the midplane, the result is\begin{align}
\label{eq:faceontotal}
    \tilde{R}_{\rm face-on}&=\begin{cases}
        \sqrt{2r_+^{-p}}\left[\frac{1-p/2}{1-v_\infty}R_{\rm LC}\right]^{1-p/2},&\gamma_\infty\sim 1
        \\
        \sqrt{2r_+^{-p}}\left[\frac{2-p}{2pr_+^{-p}}R_{\rm LC}\right]^{\frac{1-p/2}{1-p}},&\gamma_\infty \gg 1,
    \end{cases}
\end{align}
where $v_\infty$ is the terminal velocity of the flow. In the transition regime --- where $\gamma_\infty$ is of order a few --- we find that one can approximate $\tilde{R}_{\rm face-on}$ by adding the non-relativistic (NR) and force-free (FF) results in series:\begin{align}
\label{eq:seriesfaceon}
    \tilde{R}_{\rm face-on}&\approx (\tilde{R}_{\rm face-on,NR}^{-1}+\tilde{R}_{\rm face-on,FF}^{-1})^{-1}.
\end{align} 
So regardless of the jet profile, $\tilde{R}_{\rm face-on}$ scales with some inverse power of black hole spin, and the spin dependence is most pronounced at high terminal Lorentz factor. For $p=0.75$ (the default choice in this paper), we get $\tilde{R}_{\rm face-on}\propto R_{\rm LC}^{5/8}\propto a^{-5/8}$ in the non-relativistic regime and $\tilde{R}_{\rm face-on}\propto R_{\rm LC}^{5/2}\propto a^{-5/2}$ in the force-free regime. The dependence of $\tilde{R}_{\rm face-on}$ on both spin and Lorentz factor is displayed in Figure~\ref{fig:faceonscaling}, also demonstrates the validity of the approximation in Eq~\ref{eq:seriesfaceon}.

To determine the Lorentz factor for which we enter the force-free regime, we can set the two expressions in Eq.~\ref{eq:faceontotal} equal to each other and solve for $\gamma_\infty$, giving\begin{align}
    \gamma_{\rm FF,\,face-on}\gtrsim R_{\rm LC}^{\frac{p}{2(1-p)}}.
\end{align}
For $a=0.5$, we have $R_{\rm LC}\approx 24M$, indicating that we expect to enter the force-free regime when $\gamma_\infty\gtrsim 100$. 

In this manner, one can in principle measure the spins of supermassive black holes with jets viewed face-on. When a counter-jet is visible, one can look for a sharp polarimetric swing and identify it with the light cylinder (as outlined in detail in \citealp{Gelles_2025}). When a counter-jet is not visible, one can search for the impact parameter at which $\tan({\rm EVPA})=\pm 1$ and identify it with $\tilde{R}$. Combined with a measurement of the jet collimation profile, a measurement of $\tilde{R}$ can then directly constrain both spin and plasma Lorentz factor, the latter of which already has strong priors from total-intensity measurements.

We note that while we have drawn a sharp distinction between ``sources with a counter-jet" and ``sources without a counter-jet," the boundary in reality is likely far more gray. Perhaps a face-on source has a counter-jet that emits with a different EDF than the one considered here, thus rendering our binary distinction inapplicable. In cases like these, more modeling is needed, but total-intensity measurements can thankfully help constrain jet-counterjet brightness ratios to a high degree of precision.

Furthermore, while a perfectly face-on source may seem too idealized to exist in nature, several examples do actually exist. For example, the blazar PKS 1424+240 has an inclination angle less than $0.6^\circ$, and observations clearly show that the jet is seen from within its own cone \citep{kovalev_looking_2025}. The vast majority of jets in nature, however, are viewed at off-axis inclinations, and we discuss these sources in the next section. 

\section{Off-Axis Polarization}
\label{sec:offaxis}
For off-axis viewing geometries, the observer sees the emission from outside of the jet cone, causing the polarization structure to appear completely different than the results presented in the previous section. Whereas the face-on polarization structure is perfectly axisymmetric, the polarization pattern for the off-axis observer is decisively \emph{not} axisymmetric. Indeed, as discussed in \S\ref{sec:raytracing}, the shape of the jet appears to collimate in off-axis images, with observers at higher-inclinations seeing a progressively narrower projection. 

\begin{figure*}[t]
    \centering
    \includegraphics[width=0.75\textwidth]{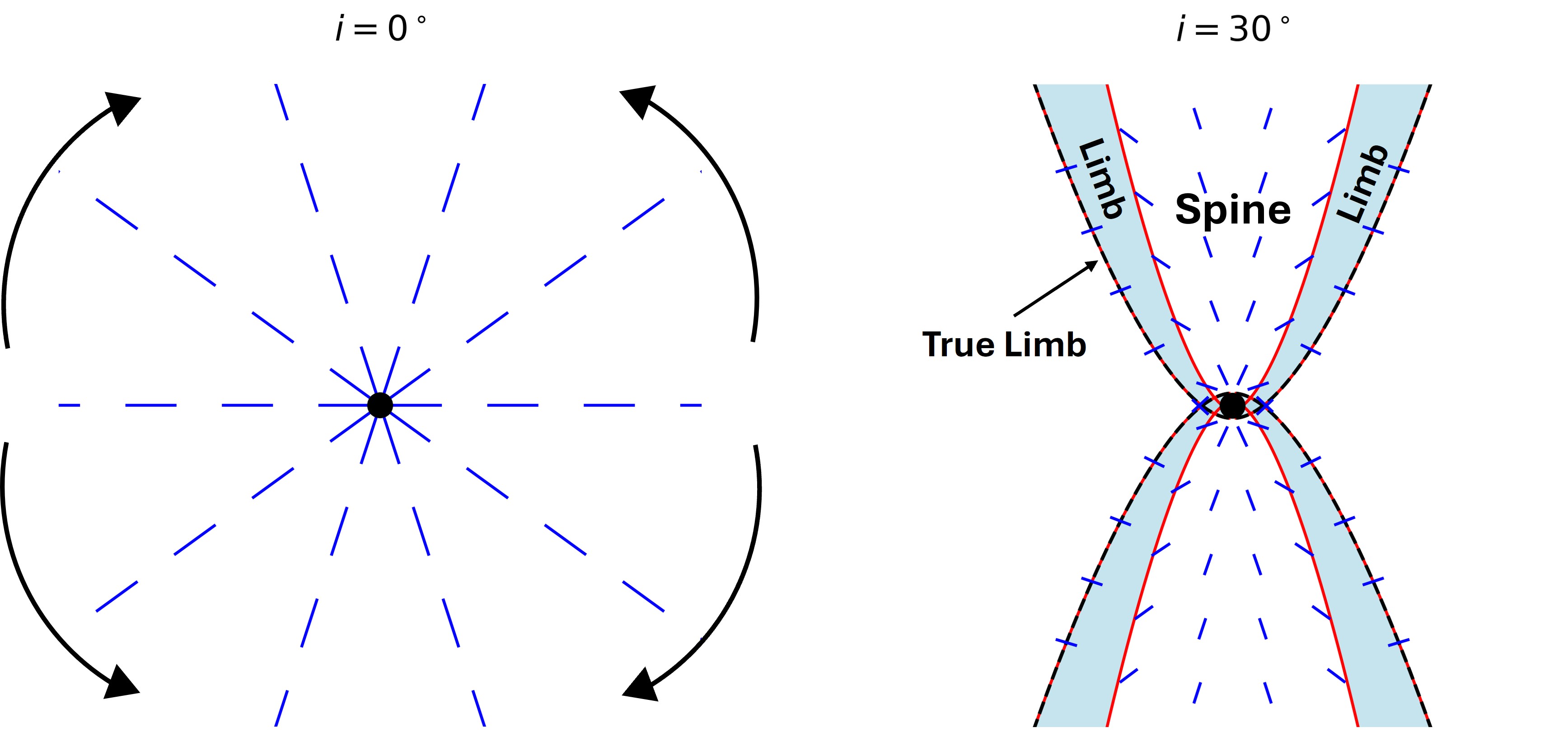}
    \caption{Appearance of a radial $(\vec{f}\propto \hat{R})$ polarization pattern as viewed from $i=0^\circ$ (left) and $i=30^\circ$ (right). In the inclined image, the foreground emission (i.e. $Y<0$) is ignored. As the inclination angle is increased, the polarization angle ``folds" up with the jet, as indicated by the black arrows. This is a purely geometric effect. The vertical axis of the image --- where polarization remains radial --- is referred to as the spine, while the regions to the sides are known as the limbs. The exact edge of the jet is the ``true limb" and is shown as a black dashed line.}
    \label{fig:foldfig}
\end{figure*}

This change in apparent jet shape has immediate implications for the polarization when viewed off-axis: as the outline of the jet ``folds up," so too will the polarization, as shown in Figure~\ref{fig:foldfig}. 

From this purely geometric ``folding" effect, we see that the off-axis polarized image should be split up into two sub-components: the \emph{spine} of the jet along the vertical axis, and the \emph{limbs} of the jet along the projected edges. In the spine of Figure~\ref{fig:foldfig}, the polarization remains radial, while in the limbs, the polarization is rotated by approximately $90^\circ$. The ``true limb" is the name we use to refer to the \emph{exact} projected of the jet, as demarcated in Figure~\ref{fig:foldfig} with a black dashed contour. 

As we will see, the polarization along the spine and the limbs each displays a characteristic swing and will hence admit a unique $\tilde{R}$ (see Eq.~\ref{eq:Rtildedef}). But the nature of the polarization swing in each sub-component of the image will be markedly different, requiring us to compute $\tilde{R}_{\rm spine}$ and $\tilde{R}_{\rm limb}$ separately. In each case, however, $\tilde{R}$ will retain strong dependence on $a$, $i$, and $\gamma_\infty$.

This spine-limb duality is commonly referenced and employed throughout the blazar literature. Observationally, a $90^\circ$ offset between the polarization of the edges and center of a jet were first noticed for the source 1055+018 \citep{attridge1999radio}, after which numerous other efforts have observed the same phenomenon in both observations and theory \citep{Pushkarev_2005,lyutikov_polarization_2005,Gabuzda_2014,Baghel_2024}. In these analyses, the presence of such a sharp distinction between spine and limb is usually attributed to a helical (i.e. $B^{\hat{\phi}}+B^Z$) magnetic field \citep{Gabuzda_2014}.

Indeed, we will show that in our model the difference between the polarization in the spine and in the limb arises mathematically because the geometry of the cross product in Eq.~\ref{eq:kcrossBfluid} is markedly different in each of these distinct regions. And when we add more complicated magnetic fields than those used to generate Figure~\ref{fig:foldfig}, we will see that the polarization pattern will actually lose its left-right symmetry across the jet. In the next two sections, we describe these effects in detail, starting with the polarization of the spine in \S\ref{sec:spinesec} and then the polarization of the limbs in \S\ref{sec:sheathsec}.

\subsection{Spine}
\label{sec:spinesec}
In this section, we discuss the spatial evolution of the polarization in the spine of the jet --- defined as the central portion of the jet with $\cos\phi\approx 0$.

To compute the EVPA along the spine, we repeat the procedure used in \S\ref{sec:polfaceon} but do not make any assumptions about the observer inclination. As with the face-on case, the Cartesian unit vectors along the spine of an off-axis jet become aligned with their cylindrical counterparts: $e_X||\pm_{\rm bg} e_{\hat{\phi}}$ and $e_Y||e_R$. So Eq.~\ref{eq:polgen} reduces to\begin{align}
    \label{eq:polspine}
    \tan({\rm EVPA})|_{\rm spine}&=\frac{f^{\hat{\phi}}}{f^R\cos i\pm_{\rm bg} f^Z\sin i},
\end{align}
where $\pm_{\rm bg}=+1$ for background ($Y>0$) emission and $\pm_{\rm bg}=-1$ for foreground ($Y<0$) emission. 

Once again, the expression in Eq.~\ref{eq:polspine} cleanly separates into a poloidal denominator with a toroidal numerator, implying that the spine polarization will swing $90^\circ$ from azimuthal to radial. 

Again substituting in the electromagnetic fields, we compute the generalization of Eq.~\ref{eq:polfaceon2}:
\begin{align}
\label{eq:aberrationspine}
    \tan({\rm EVPA})|_{\rm spine}&\approx \frac{B^R\cos i\pm_{\rm bg}B^Z\sin i}{-B^{\hat{\phi}}+\left(\frac{v^\perp}{v_{\rm FF}^\perp}\right)(E_R\cos i\pm_{\rm bg}E_Z\sin i)}.
\end{align}

Now, the numerator of Eq.~\ref{eq:aberrationspine} features two terms: $B^R\cos i$ and $B^Z\sin i$. For the face-on observer, we were easily able to toss out the $B^Z\sin i$ term. But here, we must be more careful. To see which term is dominant, we expand the electromagnetic fields (Eq.~\ref{eq:fieldexpansion}) in large $R$, finding that\begin{align}
   \frac{B^Z\sin i}{B^R\cos i}\sim \frac{Z}{R}\tan i=\frac{\tan i}{\tan \theta}.
\end{align}
Thus, $B^R$ controls the polarization of emission viewed from inside the jet cone ($i<\theta_{\rm emit}$), and $B^Z$ controls the polarization of the emission viewed from outside the jet cone $(i>\theta_{\rm emit})$. The former case corresponds to the face-on regime, while the latter case corresponds to the off-axis regime. A similar argument can be applied to show that $E^R\cos i\gg E^Z\sin i$ for the off-axis observer, and so\begin{align}
\label{eq:tanevpaoffspine}
     \tan({\rm EVPA})|_{\rm spine}&\approx \pm_{\rm bg}\frac{B^Z\sin i}{-B^{\hat{\phi}}+\frac{v^\perp}{v_{\rm FF}^\perp}E^R\cos i},
\end{align} 
where $E^R=R\Omega_FB^Z$. The above equation describes the polarization along the spine of a jet when viewed from outside of the jet cone.

\subsubsection{Role of Foreground/Background Emission}
Much like the polarization cancellation between the forward-jet and counter-jet for the face-on observer, the presence of the $\pm_{\rm bg}$ in Eq.~\ref{eq:tanevpaoffspine} allows for cancellation between the foreground and background emission observered in an off-axis setting.

This cancellation will ultimately cause the polarization along the spine to wind up more sharply, and this effect is maximized in the edge-on limit ($\cos i\to 0$). This can be seen in the Stokes basis:\begin{align}
    Q|_{\rm spine,edge-on}&=1-\frac{2(f^{\hat{\phi}})^2}{(f^{\hat{\phi}})^2+(f^{Z})^2}
    \\ U|_{\rm spine, edge-on}&=\mp_{\rm bg}\frac{2f^{\hat{\phi}}f^{Z}}{(f^{\hat{\phi}})^2+(f^{Z})^2}.
\end{align}
So if the Doppler factor and field structure of the foreground and background emission is similar, Stokes $U$ will largely cancel.

For the edge-on observer, the polarization then swings $90^\circ$ when $f^{\hat{\phi}}=f^{Z}$ and $Q$ subsequently flips sign. With no aberration (i.e. a stationary plasma), this corresponds to the point where $|B^Z|=|B^{\hat{\phi}}|$, i.e. at the light cylinder. So the cancellation between foreground and background emission can actually \emph{strengthen} the polarization swing at the light cylinder. This is completely analogous to the face-on case discussed in \S\ref{sec:polfaceon}, for which the transition from counter-jet to forward jet near the light cylinder produced a similarly strong swing.

\subsubsection{Role of Aberration}
\label{sec:aberrationsec}
When the plasma is relativistic, Stokes $Q$ will not flip sign precisely at the light cylinder, as relativistic aberration will ensure that  $f^{\hat{\phi}}$ no longer equals $f^Z$ there. This is in contrast to the face-on regime.

To understand the effects of aberration, we begin by computing the leading order velocity dependence in Eq.~\ref{eq:tanevpaoffspine}. Imposing the ``radiation condition" \citep{nathanail_black_2014} for finite Poynting flux at infinity --- which sets $\lim_{r\to\infty} E^R=B^{\hat{\phi}}$ --- we can write\begin{align}
\label{eq:evpaspinesimp}
    \tan({\rm EVPA})|_{\rm spine}\approx \mp_{\rm bg}\frac{B^Z\sin i}{B^{\hat{\phi}}(1-v_\infty \cos i)},
\end{align}
where we took $v_{\rm FF}=1$ asymptotically and $v_\infty\approx 1$ is the terminal velocity of the bulk outflow. 

Since Eq.~\ref{eq:evpaspinesimp} exhibits a polarization swing with increasing $R$, we can follow the same procedure as the face-on case (see Eqs.~\ref{eq:Rtildedef}-\ref{eq:faceonRtilde}) to quantify an effective swing location. Like the face-on polarization swing, the spine polarization also transitions gradually from azimuthal to radial, so we again find it practical to identify $\tilde{R}_{\rm spine}$ --- the ``swing location" of the off-axis jet spine --- with the point at which the EVPA has wound up exactly halfway:\begin{align}
\label{eq:spineRtilde}
    {\rm EVPA}|_{\tilde{R}_{\rm spine}}=\pm 45^\circ.
\end{align}

To compute $\tilde{R}_{\rm spine}$, we then employ that $|B^Z|=|B^{\hat{\phi}}|$ at the light cylinder and invert Eq.~\ref{eq:evpaspinesimp} to find\begin{align}
\label{eq:spineswing}
    \tilde{R}_{\rm spine}=\frac{R_{\rm LC}\sin i}{1-v_\infty\cos i}.
\end{align} 
The above expression is an increasing function of $v_\infty$, demonstrating that relativistic plasma motion causes the observed signature of the light cylinder to expand in radius. And extremizing the above expression with respect to $i$, we see that aberration along the spine is maximized when $\gamma_\infty\sin i=1$.

This is the familiar result that the apparent velocity of a relativistic source is also maximized when $\gamma_\infty\sin i=1$. In fact, we can write\begin{align}
\label{eq:appvel}
    \tilde{R}_{\rm spine}=R_{\rm LC}\frac{v_{\infty,\rm app}}{v_\infty},
\end{align}
where $v_{\rm\infty,app}$ is the apparent terminal velocity of the bulk plasma flow. Thus, it is the apparent velocity which stretches out the polarized image of the light cylinder. 

Observationally, we can then use the formula $\beta\approx Z\sin i$ together with $R\propto Z^{\frac{2}{2-p}}$ to see that the swing appears on the observer's screen at a Bardeen coordinate of \begin{align}
\label{eq:betaswingspine}
   \tilde{\beta}_{\rm spine}= \beta_{\rm LC}\left(\frac{v_{\infty,\rm app}}{v_\infty}\right)^{1-p/2}.
\end{align}

\begin{figure*}[t]
    \centering
    \includegraphics[width=\textwidth]{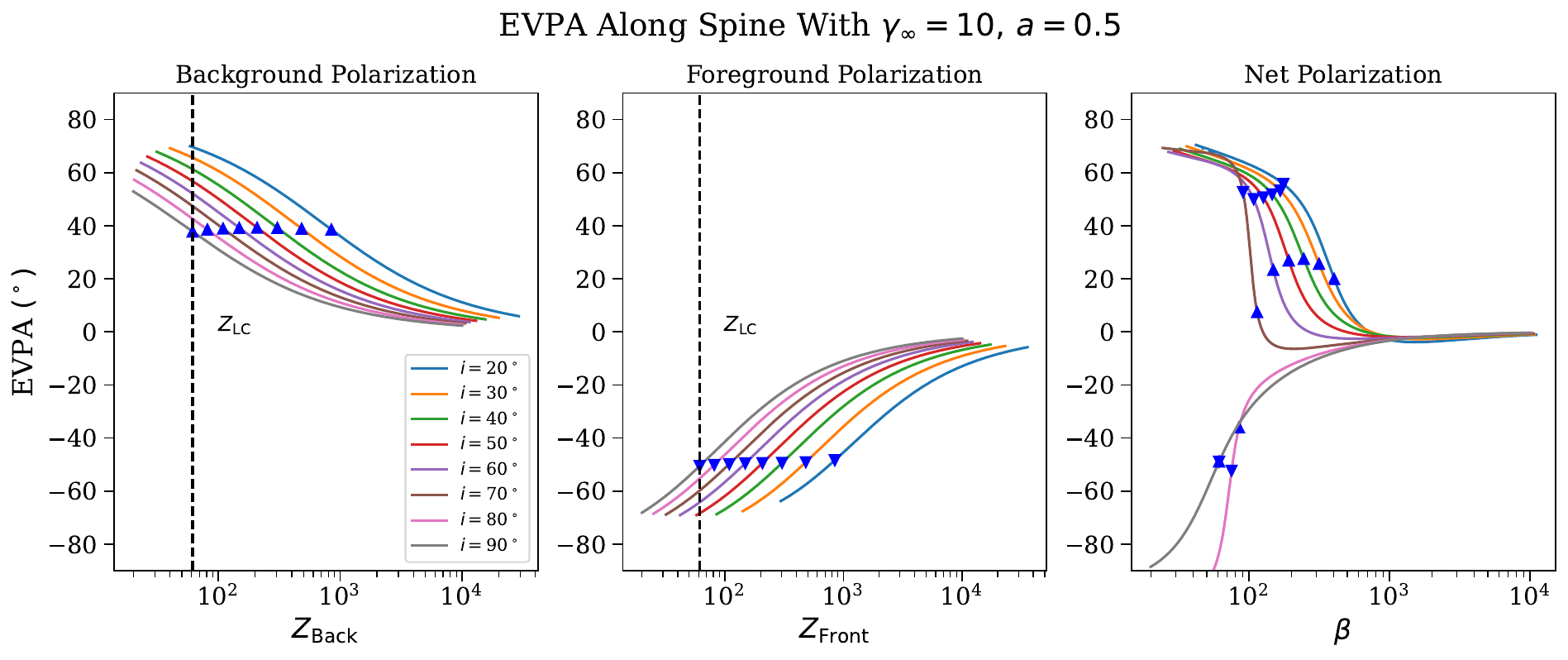}
    \caption{Spine polarization as a function of inclination. The individual panels represent the background emission (left), foreground emission (middle), and net sum (right). The location of the true light cylinder is given by a black dashed line, and the predictions of Eq.~\ref{eq:spineswing} for $\tilde{R}_{\rm spine}$ (the location where ${\rm EVPA}=\pm 45^\circ$) are plotted as face-up triangles for the background emission and face-down triangles for the foreground emission. In all cases, we see that Eq.~\ref{eq:spineswing} is an excellent prediction for the location of the polarization swing. As inclination is increased, the location of the polarization swing begins to align with the true light cylinder. On the curves of the right panel, we include both the face-up and face-down triangles, respectively giving the predictions for both the background and foreground swing locations.  The location where the net ${\rm EVPA} = 45^\circ$ lies roughly between these two estimates.}
    \label{fig:spinecompareinc}
\end{figure*}

Unlike the face-on case, the mapping between $\tilde{R}$ and $\tilde{\beta}$ for the off-axis spine will differ between the foreground and background: the background emission will attain a polarization angle of $45^\circ$ at a distinct impact parameter from that of the foreground emission. This distinction is displayed in Figure~\ref{fig:spinecompareinc}, for which we plot the background spine polarization, foreground spine polarization, and net spine polarization for different observer inclinations. In the plots of the foreground and background, we attach a single blue marker (face-up triangle for background, face-down triangle for foreground) to each polarization curve, corresponding to the predicted emission radius given by Eq.~\ref{eq:spineswing}. In the plot of the net polarization, we show \emph{both} blue markers. These two markers roughly bound the ``swing" region, where the net polarization changes most rapidly from azimuthal to radial.

\begin{figure}[h]
    \centering
    \includegraphics[width=0.4\textwidth]{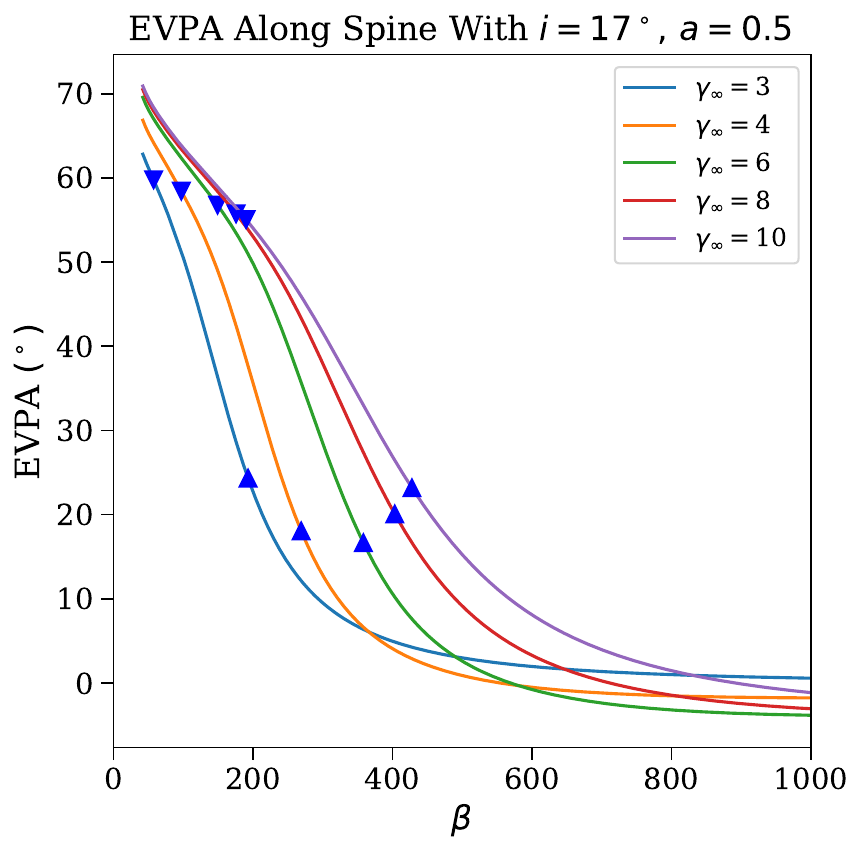}
    \caption{Net spine polarization as a function of $\gamma_\infty$. Increasing $\gamma_\infty$ generically makes the location of the polarization swing move farther out. As with Figure~\ref{fig:spinecompareinc}, each curve has two markers in blue: face-up triangles denoting the prediction (Eq.~\ref{eq:spineswing}) for the effective swing location along the background emission, and face-down triangles denoting the prediction for the effective swing location along the foreground emission. The location where the net EVPA$=45^\circ$ lies roughly between these two estimates of the swing location.}
    \label{fig:spinecomparegamma}
\end{figure}

\begin{figure}[h]
    \centering
    \includegraphics[width=0.4\textwidth]{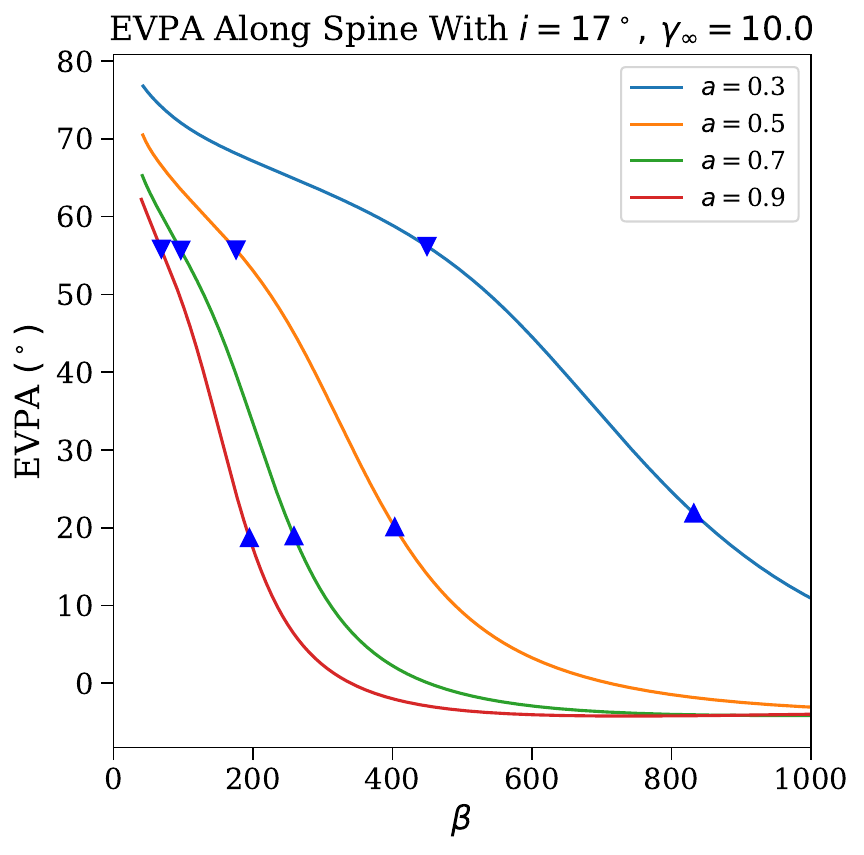}
    \caption{Net spine polarization as a function of $a$. Increasing $a$ generically makes the location of the polarization swing move farther in, thus providing an useful diagnostic of spin. As with Figure~\ref{fig:spinecompareinc}, each curve has two markers in blue: face-up triangles denoting the prediction (Eq.~\ref{eq:spineswing}) for the effective swing location along the background emission, and face-down triangles denoting the prediction for the effective swing location along the foreground emission. The location where the net EVPA$=45^\circ$ lies roughly between these two estimates of the swing location.}
    \label{fig:spinecomparespin}
\end{figure}

In Figure~\ref{fig:spinecomparegamma} and \ref{fig:spinecomparespin}, we repeat these plots of the EVPA for different Lorentz factors and spins respectively. In these two plots, we plot only the net polarization but again include the two blue markers corresponding to the swing location of both background and foreground.

Now, Eq.~\ref{eq:betaswingspine} is an extremely convenient relationship because $v_{\infty}\approx 1$ for any relativistic source, and $v_{\infty,\rm app}$ is \emph{directly} observable from the data \citep{rees1966appearance}. Thus, $R_{\rm LC}$ can be inferred very cleanly from a measurement of a polarization swing at $\tilde{\beta}_{\rm spine}$. And with knowledge of the light cylinder radius, one can then constrain spin, as $R_{\rm LC}\propto a^{-1}$. This spin dependence is manifest in Figure~\ref{fig:spinecomparespin}.

We note that the results of spine polarization here appear to be relatively insensitive to the transverse location of emission within the spine. While Eqs.~\ref{eq:spineswing} and \ref{eq:betaswingspine} were computed for $\cos\phi=0$ exactly, the result holds for all values of $\phi$ away from $\pm\pi/2$, as we will eventually show in Figure~\ref{fig:evpavsphi}. The small range of emission locations with $\phi\approx \pm\pi/2$ instead constitute a narrow portion of the projected jet that we identify with the limbs. This portion of the jet image is discussed below.

\subsection{Limbs}
\label{sec:sheathsec}
In this section, we discuss the spatial evolution of the polarization in the limbs --- defined as the region near the projected edges of the jet (i.e. with $\sin\phi\approx 0$). While this definition of the limbs is somewhat vague, it will become clear that the limbs are easily distinguished from the spine via polarization. Importantly, the limbs are present only for observers who are \emph{not} in the face-on regime; portions of the jet viewed from within the jet cone do not have a projected edge at all. 

In the limbs, the Cartesian coordinates again align with their cylindrical counterparts, but in a manner different from before: $e_X||e_R$ and $e_Y||e_{\hat{\phi}}$. So we can write\begin{align}
\label{eq:polsheath}
    \tan({\rm EVPA})|_{\rm limb}&=-\frac{f^R}{ f^{\hat{\phi}}\cos i\pm_R f^Z\sin i},
\end{align}
where $\pm_R=+1$ on the right side of the image $(X>0)$ and $\pm_R=-1$ on the left side of the image $(X<0)$. 

Comparing Eq.~\ref{eq:polsheath} to Eq.~\ref{eq:polspine}, we see a dramatically different polarization structure along the limbs than along the spine; as the magnetic field winds up, $f^R$ becomes the dominant component of the polarization, causing an asymptotically \emph{azimuthal} polarization pattern along the limbs. This is exactly the ``folding up" of the polarization that was introduced in Figure~\ref{fig:foldfig}. This folding of the polarization constitutes the crux of the spine-limb duality.

\subsubsection{Role of Aberration}
Like the spine, aberration plays a meaningful role in controlling the asymptotic polarization structure of the limbs. In particular, if the plasma is relativistic, then the numerator of Eq.~\ref{eq:polsheath} can actually cross zero, causing a $180^\circ$ swing in polarization. 

To see this, we turn again to Eq.~\ref{eq:polcross} to find\begin{align}
    f^R|_{\rm limb}&=(\vec{k}\times\vec{B})^R+\left(\frac{v^\perp}{v_{\rm FF}^\perp}\right)E^R
    \\ \label{eq:fRsheathbefore}&=-(B^{\hat{\phi}}\cos i\pm_RB^Z\sin i)+\frac{v^\perp}{v_{\rm FF}^\perp}E_R
    \\ &
    \approx
     B^{\hat{\phi}}(v_\infty-\cos i)\mp_R B^Z\sin i\label{eq:fRapprox}
\end{align}
where in the last line we applied the radiation condition and set $v_{\rm FF}^\perp=1$.

Now, if the plasma is traveling in the $+\hat{\phi}$ direction, then MHD theory tells us that $B^{\hat{\phi}}<0$ with $B^Z>0$. Therefore, $f^R|_{\rm limb}$ can cross zero along the receding (right) edge of the jet when $v_\infty<\cos i$, and $f^R|_{\rm limb}$ can cross zero along the approaching (left) edge of the jet when $v_\infty>\cos i$. In other words,
\begin{align}
\label{eq:gammafliplimb}
    \gamma_\infty\sin i<1\Longrightarrow \text{pol. swing on receding edge}
    \\
   \nonumber \gamma_\infty\sin i>1\Longrightarrow \text{pol. swing on boosted edge}.
\end{align}
The polarized image of the limbs is therefore strongly asymmetric; only one side of the image has a polarization swing, depending on the magnitude of $\gamma_\infty\sin i$.

This result has immediate observational consequences. If one measures a polarization swing on one side of the image and identifies it with the receding/approaching side (via Doppler arguments), then one can constrain the sign of $\gamma_\infty\sin i-1$. Or conversely, if $\gamma\sin i$ is known, then it can be combined with a measurement of polarization symmetry to constrain direction of rotation inside the jet. 

As with the spine, we can also learn about the spin of the black hole by quantifying the location of the polarization swing along the limbs. Except unlike the case of the spine, the limb polarization swings when $f^R=0$. Hence we identify $\tilde{R}_{\rm limb}$ --- the ``swing location" along the limbs ---  with the point where the EVPA is zero:\begin{align}
\label{eq:Rtildelimb}
   {\rm EVPA}|_{\tilde{R}_{\rm limb}}=0,
\end{align}
so that the EVPA at $\tilde{R}_{\rm limb}$ is offset from that of $\tilde{R}_{\rm spine}$ by $45^\circ$ (cf. Eq.~\ref{eq:spineRtilde}).

Inverting Eq.~\ref{eq:fRapprox} to leading order in $R/R_{\rm LC}$, we then find\begin{align}
\label{eq:Rsheath}
    \tilde{R}_{\rm limb,0}=\frac{R_{\rm LC}\sin i}{|\cos i-v_\infty|},
\end{align}
where the ``0" subscript indicates that the equation is accurate to leading order in large $R$. Once again, we have recovered a linear dependence in the light cylinder radius.

This expression for $\tilde{R}_{\rm limb,0}$ can be simplified even further using the Lorentz transformation of angles\begin{align}
\label{eq:Rswingsheathvapp}
    \tilde{R}_{\rm limb,0}&=\frac{\tilde{R}_{\rm spine}}{|\cos i'|}=R_{\rm LC}\left|\frac{v_{\infty,\rm app}}{v_\infty\cos i'}\right|,
\end{align}
where\begin{align}
    \cos i'=\frac{\cos i-v_\infty}{1-v_\infty\cos i}
\end{align}
is the observer inclination measured in the fluid frame. The Bardeen coordinate of the swing is then given by\begin{align}
    \label{eq:betaswingsheath}
    \tilde{\beta}_{\rm \,limb,0}= \beta_{\rm LC}\left|\frac{v_{\infty,\rm app}}{v_\infty\cos i'}\right|^{1-p/2}.
\end{align}

The factor of $\cos i'$ is what differentiates the limbs from the spine, and it comes purely from geometry. Along the spine, the emitted wavevector has no component tangent to $B^{\hat{\phi}}$ (the dominant component of the magnetic field). But along the limbs, the wavevector does pick up a component parallel to $B^{\hat{\phi}}$. So in the fluid frame, where $\vec{f}=\vec{k}\times\vec{B}$, we have \begin{align}
    |\vec{k}\times \vec{B}|_{\rm spine}\sim |k_\perp B^{\hat{\phi}}|&=|B^{\hat{\phi}}|,
    \\
    |\vec{k}\times \vec{B}|_{\rm limb}\sim |k_\perp B^{\hat{\phi}}|&=|B^{\hat{\phi}}\cos i'|.
\end{align}

\begin{figure}[h]
    \centering
    \includegraphics[width=0.5\textwidth]{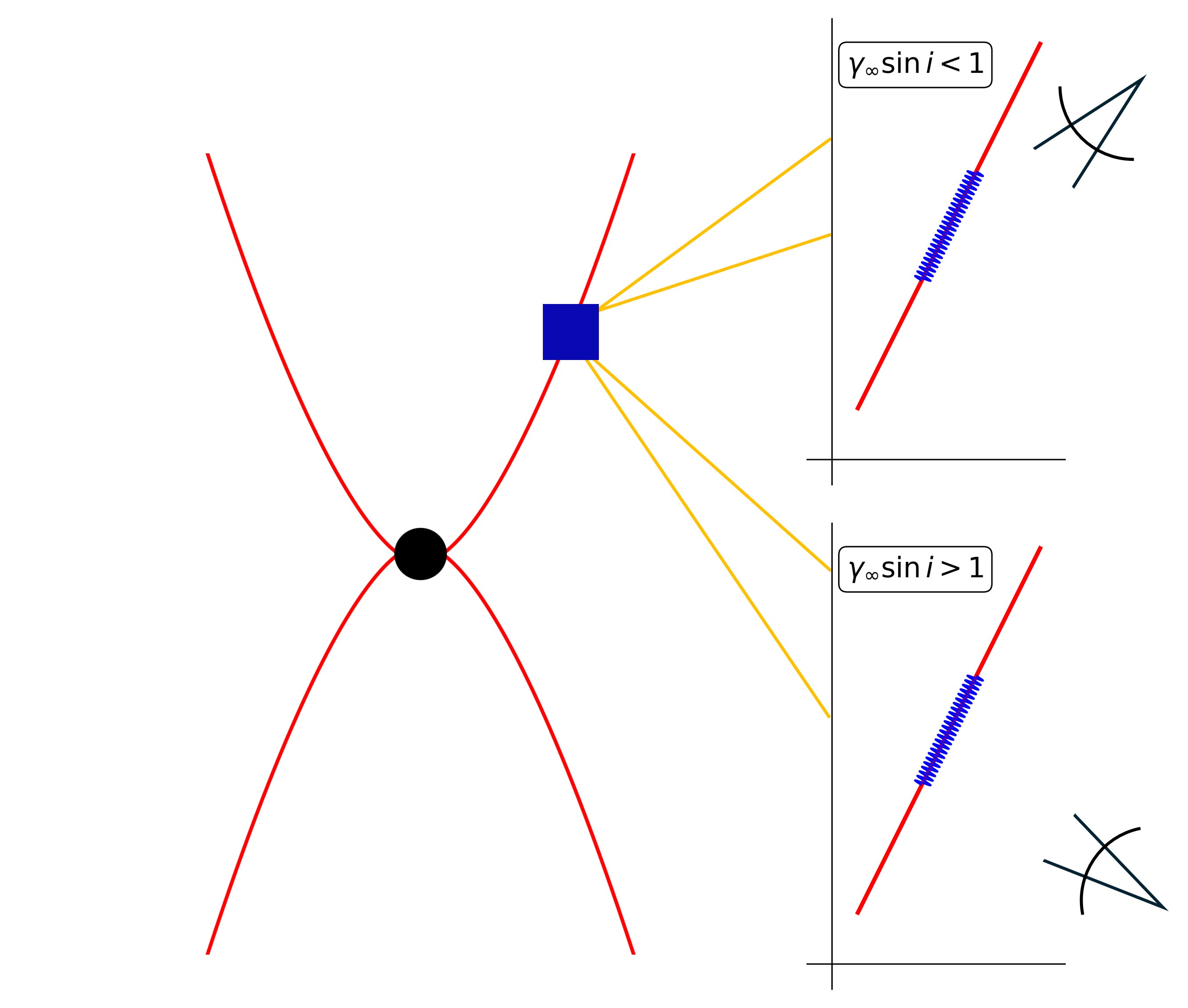}
    \caption{Boosting into the local frame of a fluid parcel (shown as a square, with magnetic fieldline in red and electron gyromotion in blue). In this fluid frame, the orientation of the observer's line of sight depends on the magnitude of $\gamma_\infty\sin i$. If $\gamma_\infty\sin i>1$, then the observer lies above the emitting material. If $\gamma_\infty\sin i<1$, then the observer lies below the emitting material. This geometric flip causes the polarization pattern to reflect across the image when $\gamma_\infty\sin i$ crosses 1, as shown in Figures~\ref{fig:sheathcomparegamma} and \ref{fig:semipolmaps}.}
    \label{fig:inclinationfig}
\end{figure}

Furthermore, the factor of $\cos i'$ in the denominator of Eq.~\ref{eq:Rswingsheathvapp} provides an intuitive reason for why the location of the polarization swing can switch from the receding limb to the boosted limb of the jet. For $\gamma_\infty\sin i<1$, the observer is located above the midplane $(\cos i'>0)$ in the fluid frame. But for $\gamma_\infty\sin i>1$, the observer is located below the midplane $(\cos i'<0)$ in the fluid frame. Therefore, as $\gamma_\infty\sin i$ is increased past 1, the polarization pattern will \emph{reflect} across the vertical axis of the image. This is depicted pictorially in Figure~\ref{fig:inclinationfig}.

Finally, we remark that while Eq.~\ref{eq:Rsheath} captures the leading order spin-dependence of the swing location, higher order corrections can be necessary to overcome the two approximations made in Eq.~\ref{eq:fRapprox} (that the radiation condition holds exactly and that $v^\perp/v_{\rm FF}^\perp=v_\infty$).

To this end, we derive the first perturbative corrections to Eq.~\ref{eq:Rsheath}, finding\begin{align}
\label{eq:Rsheathpert}
    \tilde{R}_{\rm limb}&=\tilde{R}_0+\mathcal{A}\tilde{R}_0^{\frac{2-3p}{2-p}}+\mathcal{B}\tilde{R}_0^{-1}+\mathcal{O}(\tilde{R}_0^{-\frac{4p}{2-p}}),
\end{align}
where $\mathcal{A}$ and $\mathcal{B}$ are computed explicitly in Appendix~\ref{app:pertsheath}. The interpretation here is that $\mathcal{A}$ encodes the dependence on the jet geometry, and $\mathcal{B}$ encodes a higher-order correction in spin.
\begin{figure*}[t]
    \centering
    \includegraphics[width=\textwidth]{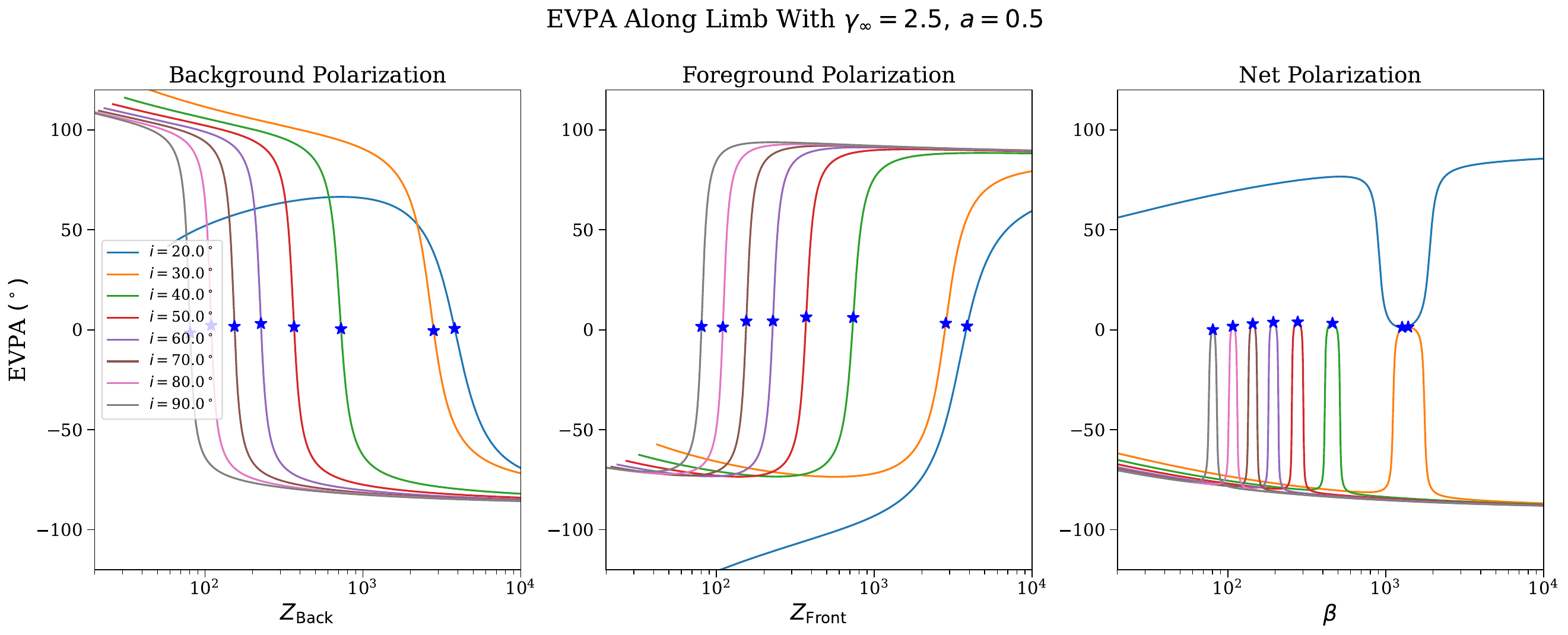}
    \caption{Limb polarization as a function of inclination, plotted for $\phi-\phi_{\rm TL}=1^\circ$, where $\phi$ is the background azimuthal coordinate of the emission and $\phi_{\rm TL}$ is the azimuthal coordinate of the true limb (see \S\ref{sec:truelimb} for more detail). The individual panels represent the background emission (left), foreground emission (middle), and net sum (right). In all three cases, the predicted location of $\tilde{R}_{\rm limb}$ (Eq.~\ref{eq:Rsheathpert}; the location at which $f^R=0$) is marked as a blue star. Unlike the spine (Figure~\ref{fig:spinecompareinc}), only a single blue star is needed here because the foreground and background coincide at the limb. The polarization swings rapidly around $\tilde{R}_{\rm limb}$ in all cases. Note that we have plotted only one of the two limbs here: the receding limb for $i=20^\circ$ and the approaching limb for all other inclinations (see Eq.~\ref{eq:gammafliplimb}). }
    \label{fig:sheathcompareinc}
\end{figure*}
\begin{figure*}[t]
    \centering
    \includegraphics[width=\textwidth]{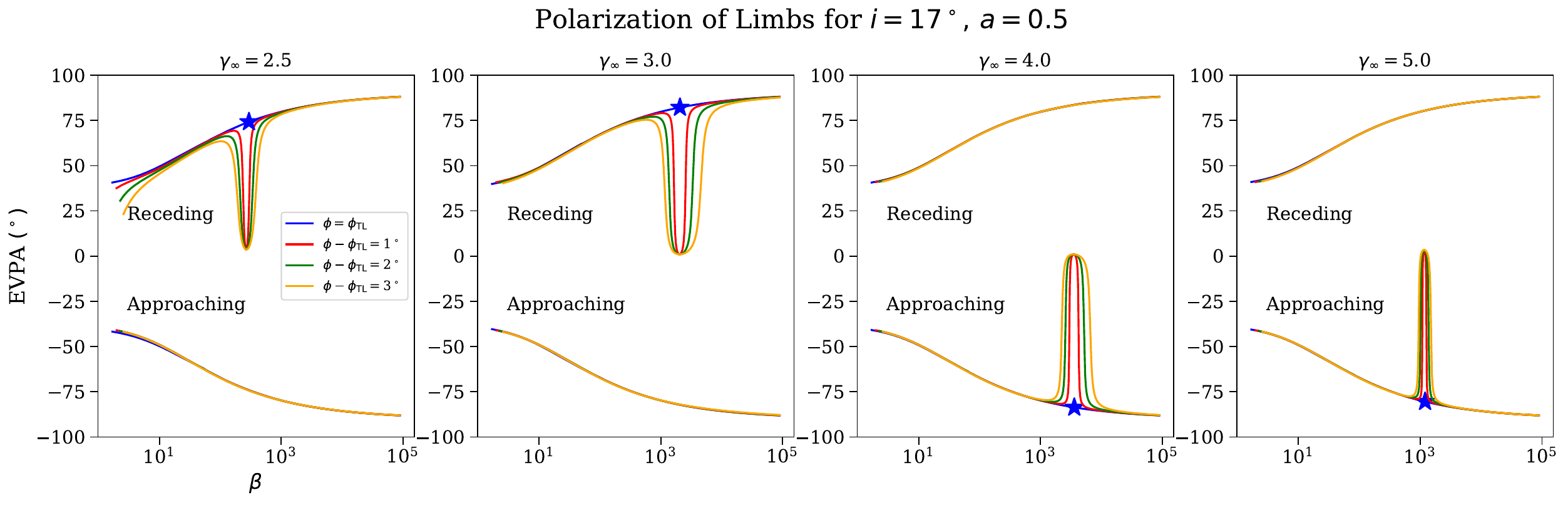}
    \caption{Polarization of the limb, plotted for multiple different values of $\gamma_\infty$ and $\phi$. The angles are measured relative to the ``true limb" $\phi_{\rm TL}$, which denotes the genuine projected edge of the jet. The blue stars denote the prediction of Eq.~\ref{eq:Rsheathpert} for the location of $\tilde{R}_{\rm limb}$ and hence $\tilde{\beta}_{\rm limb}$ (the point at which $f^R=0$). For $i=17^\circ$, we have $\csc i\approx 3.42$, which is why the swing switches from the receding edge to the boosted edge after the second panel.}
    \label{fig:sheathcomparegamma}
\end{figure*}

The limb polarization is plotted in Figure~\ref{fig:sheathcompareinc} for several different inclination angles. Like we did for the spine, we separate the contributions from foreground and background emission before combining them. We see that regardless of inclination angle, the foreground EVPA and the background EVPA each swing $180^\circ$, passing through a purely radial polarization angle precisely when $R=\tilde{R}_{\rm limb}$. When the foreground and background combine, they yield a polarization pattern which appears to rise $90^\circ$ and then fall $90^\circ$. This stands in contrast to the swing along the spine, for which the polarization angle only swings $90^\circ$ in one direction. 

Furthermore, the limb polarization curves differ from those of the spine because $\tilde{R}_{\rm limb}$ maps onto the (approximately) same image coordinate for both foreground and background emission; the ray-tracing from $\tilde{R}_{\rm limb}\to \tilde{\beta}_{\rm limb}$ is unique. For this reason, we mark the location of the limb swing in Figure~\ref{fig:sheathcompareinc} with a single blue star, in contrast to the separate foreground and background markers used in Figure~\ref{fig:spinecompareinc}.

We stress that our model is optimized to accurately predict EVPA --- not total intensity. In practice, therefore, the exact ways in which foreground and background combine may be somewhat different than the right panel of Figure~\ref{fig:sheathcompareinc} and may depend on the observing kernel. However, the existence of a net $180^\circ$ polarization swing is robust and model-agnostic; regardless of the emissivity profile, both foreground and background must pass through $f_R=0$ at $\tilde{R}_{\rm limb}$, producing some kind of swing there.

In Figure~\ref{fig:sheathcomparegamma}, we then plot the limb polarization along contours of $\phi$ for several different terminal Lorentz factors. When $i=17^\circ$ is fixed, then the location of the polarization swing should switch from one limb to the other when  $\gamma_\infty=\csc i=3.42$. Thus, the left two panels of the figure (for which $\gamma_\infty=2.5$ and $\gamma_\infty=3$) show the polarization swing along the receding edge of the jet, while the right two panels of the figure (for which $\gamma_\infty=4$ and $\gamma_\infty=5$)  show the polarization swing along the boosted edge of the jet.  In all cases, Eq.~\ref{eq:Rsheathpert} remains an excellent approximation to the polarization swing location along the limbs. 

Figure~\ref{fig:sheathcomparegamma} additionally reveals that the shape of the limb polarization swing is not uniform across $\phi$. In fact, as one approaches the ``true limb" --- or genuine edge --- of the jet, the swing disappears all together. We discuss this phenomenon below.

\subsubsection{True Limb}
\label{sec:truelimb}
The true limb (TL) of the jet is defined as the observed boundary of the limb: it is the \emph{exact} projected edge of the jet. Along the true limb, the wave-vector is tangent to the jet wall:\begin{align}
\label{eq:TLcondition}
    \vec{k}\cdot\nabla\psi|_{\rm TL}=0,
\end{align}
which is equivalent to the condition that each geodesic intersect the true limb exactly once. While the location of the true limb always satisfies the approximate relation $\sin\phi\approx 0$ for inclined observers (with the approximation becoming more accurate farther down the jet), exact equality need not hold. The true limb is depicted pictorially in Figure~\ref{fig:foldfig}. 

The TL of a single fieldline occupies a measure-zero subset of the emission volume. However, the TL displays a number of curious properties that we expect to become observable in astrophysical scenarios, where emission is summed along a bundle of fieldlines (each of which has its own TL). In particular, the polarization swing \emph{completely disappears} along the true limb. This is because when the numerator of Eq.~\ref{eq:polsheath} vanishes on the true limb, the denominator also vanishes:
\begin{align}
    f|_{\rm TL}=0\quad {\rm at}\quad R=\tilde{R}_{\rm limb}.
\end{align}
The proof is given in Appendix~\ref{app:truelimb}.

This explains the appearance of the polarization swings in Figure~\ref{fig:sheathcomparegamma}. Along any contour of fixed $\phi$, a polarization swing emerges that is localized around $\tilde{R}_{\rm limb}$. But as $\phi$ approaches $\phi_{\rm TL}$ (the azimuthal coordinate of the true limb), the $180^\circ$ polarization swing gets progressively sharper and sharper. At the exact location of the true limb, the polarization swing becomes infinitely sharp and disappears completely.

\begin{figure*}[t]
    \centering
    \includegraphics[width=0.9\textwidth]{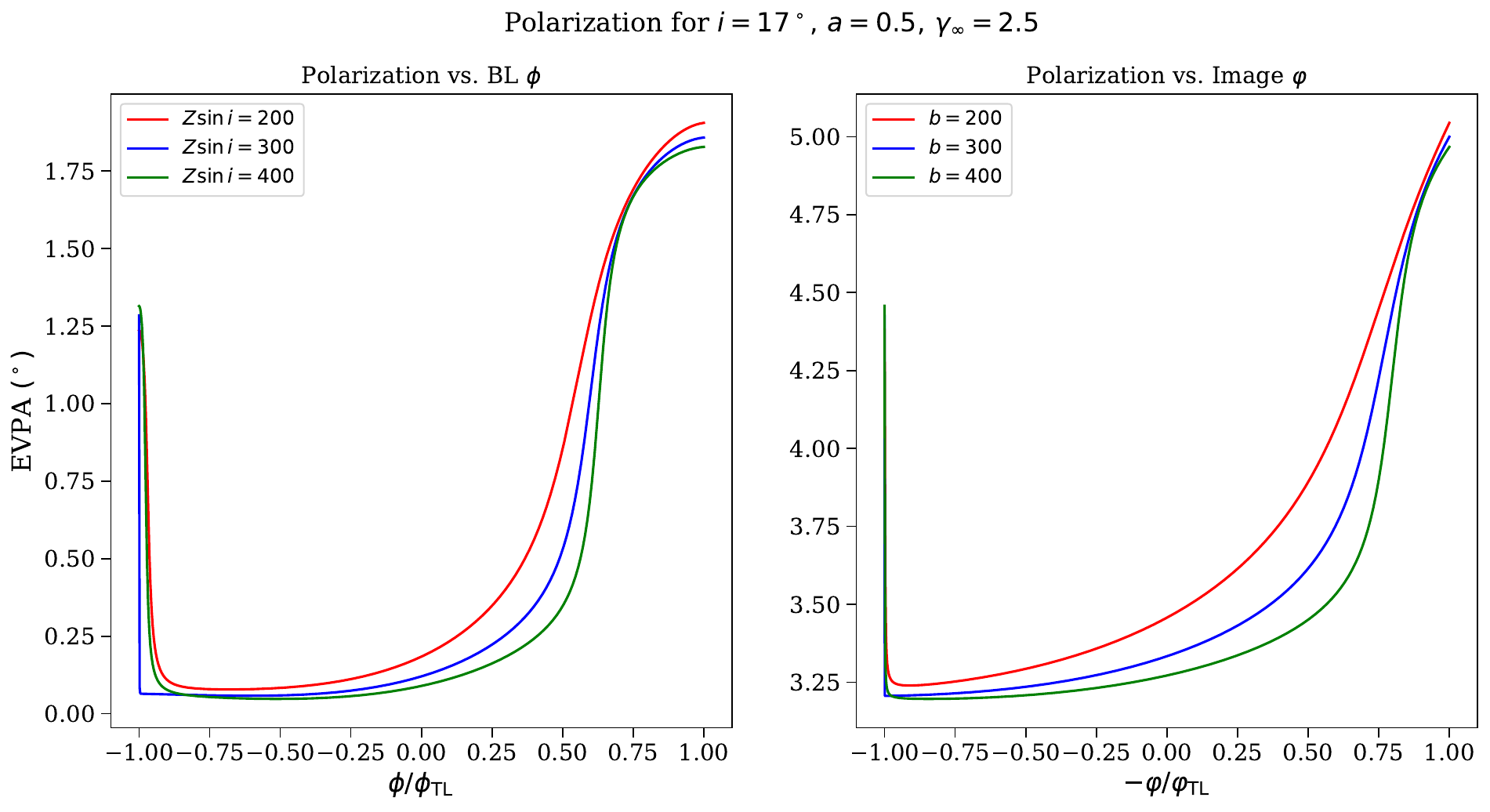}
    \caption{Left: EVPA as a function of the emission coordinate $\phi$ along a contour of fixed $Z$ (with both coordinates referring to the location of the background emission). Right: The observable analog; EVPA as a function of the image coordinate $\varphi$ along a fixed impact parameter $b$. In both cases, the curve approaches a step function at the singular point of the true limb. Searching for this behavior could shed clues on black hole spin, as the impact parameter for which the EVPA develops this kink depends sensitively on spin.}
    \label{fig:evpavsphi}
\end{figure*}

The transverse structure of the polarization along the jet is thus incredibly rich. In Figure~\ref{fig:evpavsphi}, we plot the transverse EVPA as a function of both $\phi$ and $\varphi$. In both cases, the transverse structure develops a sharp kink as one moves farther away from the black hole, and at $\tilde{R}_{\rm limb}$, the swing becomes a step function at the true limb. 

As a result, one can actually identify $\tilde{R}_{\rm limb}$ in two different ways: (a) by finding the impact parameter at which the limb EVPA swings through $0^\circ$, or (b) by finding the impact parameter at which the transverse EVPA structure develops a kink, i.e., sharp angular feature. The latter method method makes a strong case for high angular resolution in future interferometric upgrades, as identifying a kink in a polarized image requires high fidelity.

\begin{figure*}[t]
    \centering
    \includegraphics[width=\textwidth]{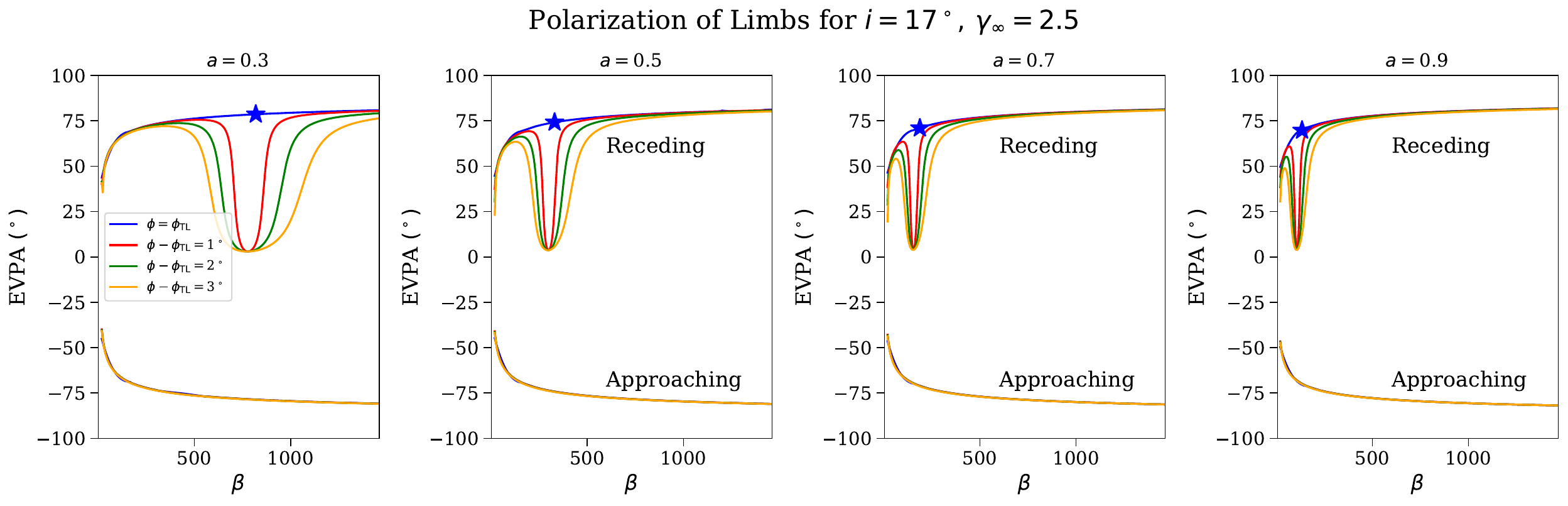}
    \caption{Polarization of the limb, plotted for multiple different values of $a$ and $\phi$. The terminal Lorentz factor is fixed at $\gamma_\infty=2.5$. In all cases, we have $\gamma_\infty<\csc i$, so the swing appears on the receding edge. The blue stars denote the prediction of Eq.~\ref{eq:Rsheathpert} for the location of $\tilde{R}_{\rm limb}$ and hence $\tilde{\beta}_{\rm limb}$ (the point at which $f^R=0$).}
    \label{fig:sheathcomparespin}
\end{figure*}

Regardless of how $\tilde{R}_{\rm limb}$ is identified, it can then be used to constrain spin, as $\tilde{R}_{\rm limb}\propto R_{\rm LC}\propto a^{-1}$. Indeed, one can use Eq.~\ref{eq:Rsheath} to jointly constrain black hole spin and plasma Lorentz factor directly from a measurement of $\tilde{R}$. The spin dependence of the limb polarization swing is demonstrated in Figure~\ref{fig:sheathcomparespin}.


We note that the signatures discussed above apply only to the forward jet --- which is most easily seen due to Doppler enhancement. However, there are still important polarization diagnostics in the off-axis counter-jet, to the extent that it can be seen. We discuss these signatures next.

\subsection{Off-Axis Counter-Jet}
\label{sec:offaxisCJsec}
In off-axis images, the counter-jet lies in the bottom half ($\beta<0$) of the image, in contrast to the forward jet in the top half $(\beta>0)$ of the image. Since the counter-jet is boosted \emph{away} from the observer, most of the aberration results previously derived will drop out, and the counter-jet polarization pattern will reduce back to the non-relativistic regime.

To see this, observe how the field solutions to axisymmetric GRMHD transform under reflection about the midplane:\begin{align}
    B^R(R,\phi,Z)&=-B^R(R,\phi,-Z)
    \\
    B^Z(R,\phi,Z)&=B^Z(R,\phi,-Z)
    \\
    B^{\hat{\phi}}(R,\phi,Z)&=-B^{\hat{\phi}}(R,\phi,-Z)
    \\
    E^R(R,\phi,Z)&=B^R(R,\phi,-Z)
    \\
    E^Z(R,\phi,Z)&=-E^Z(R,\phi,-Z)
    \\
    E^{\hat{\phi}}(R,\phi,Z)&=0.
\end{align} 
Therefore, the radiation condition \citep{nathanail_black_2014} in the southern hemisphere picks up a minus sign:\begin{align}
    \lim_{r\to\infty} E^R|_{Z<0}=-B^{\hat{\phi}}.
\end{align}
As a result, all of the factors of $1-v$ that led $\tilde{R}$ to grow large in the forward jet will become $1+v$ in the counter-jet, which is bounded from below. Indeed, we can define $\tilde{R}_{\rm spine,CJ}$ --- the ``swing location" of the counter-jet spine --- in the analogous manner to the forward jet spine:\begin{align}
    {\rm EVPA}|_{\tilde{R}_{\rm spine,CJ}}&=\pm 45^\circ.
\end{align}
Then, we can compute 
\begin{align}
\label{eq:CJpred}
    \tilde{R}_{\rm spine,CJ}&=\frac{R_{\rm LC}\sin i}{1+v_\infty\cos i}\approx R_{\rm LC}\tan\left(\frac{i}{2}\right),
\end{align}
where the approximation took $v_\infty\approx 1$. The factor of $\tan(i/2)$ means that the swing in the counter-jet is only visible for high inclinations.

Meanwhile, in the limbs, we can repeat the calculation of Eq.~\ref{eq:fRapprox} but with a flipped sign in the radiation condition. From this, we see that $f^R$ can now flip sign in the counter-jet \emph{only} on the approaching edge. This means that along the approaching edge of the counter-jet, we can define a swing at a location analogous to the limb of the forward jet:\begin{align}
    {\rm EVPA}|_{\tilde{R}_{\rm limb,CJ}}&=0^\circ.
\end{align}
This swing occurs at \begin{align}
    \tilde{R}_{\rm limb,CJ}=\frac{R_{\rm LC}\sin i}{\cos i+v_\infty}\approx R_{\rm LC}\tan\left(\frac{i}{2}\right).
\end{align}
Therefore, the counter-jet limb swings at roughly the same place as the counter-jet spine. Furthermore, $\tilde{R}_{\rm limb,CJ}$ and $\tilde{R}_{\rm spine,CJ}$ both fall \emph{within} the light-cylinder, ensuring that the emission will not be significantly de-boosted and should hence be visible with sufficiently high-resolution imaging.

\begin{figure*}[t]
    \centering
    \includegraphics[width=\textwidth]{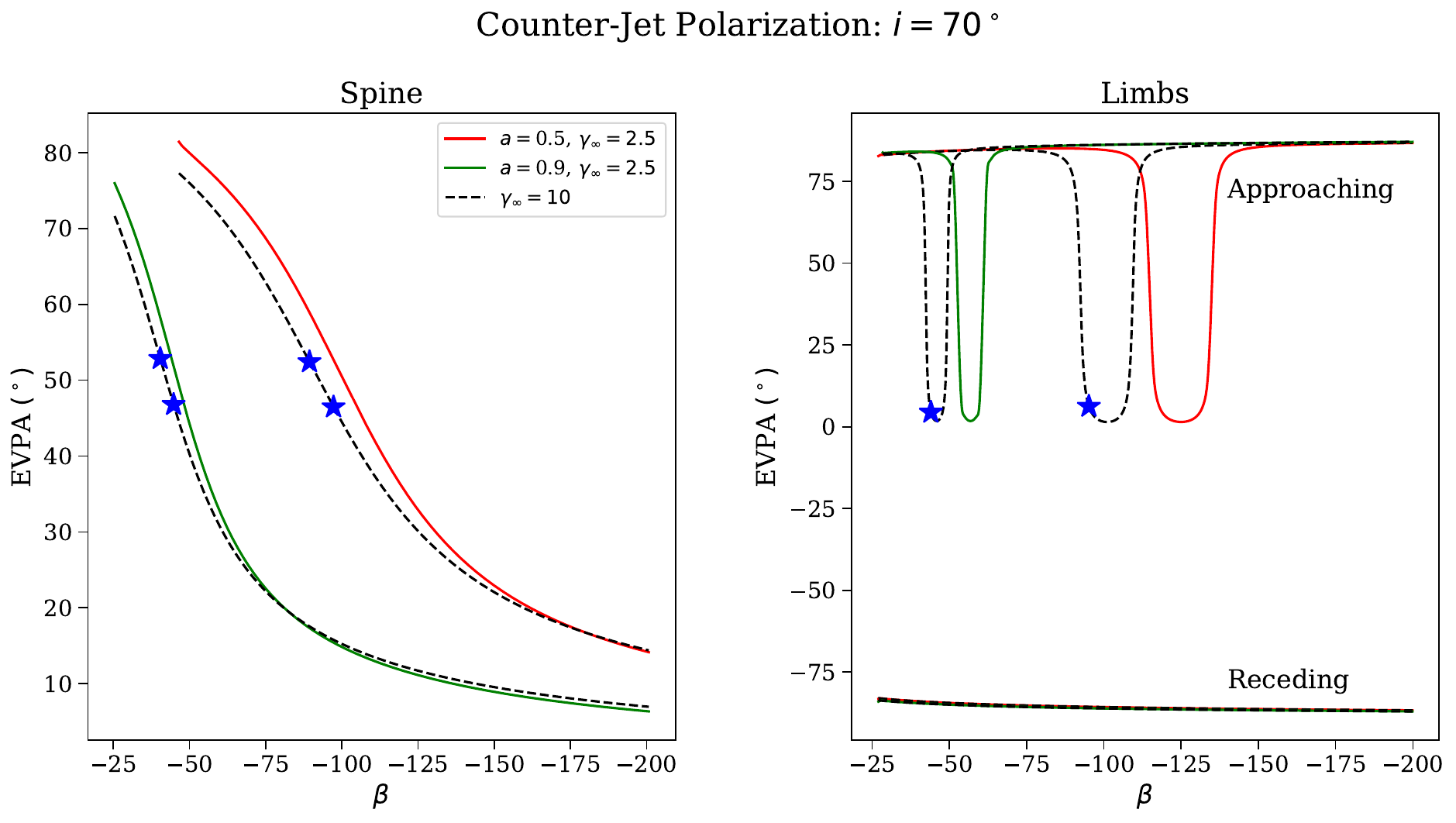}
    \caption{Polarization along the spine (left) and limbs (right) of the counter-jet when viewed at high inclination. For a single spin, both the spine EVPA and the limb EVPA swing at roughly the same location; the prediction for this swing location (Eq.~\ref{eq:CJpred}) is most accurate in the relativistic limit and is shown as a blue star. In the spine, the $\gamma_\infty$ dependence is very weak, and in the limbs, the $\gamma_\infty$ dependence is modest. The fieldline used for this plot is $\psi=0.15$.}
    \label{fig:offaxisCJ}
\end{figure*}

We plot the polarization along the spine and limbs of the counter-jet in Figure~\ref{fig:offaxisCJ}. The polarization is plotted at a high inclination $(i=70^\circ)$ and with a particularly collimated fieldline ($\psi=0.15$), as this leads to the clearest swings. In the spine, there is minimal dependence on $\gamma_\infty$. In the limb, there is some dependence on $\gamma_\infty$, but it is still much weaker than the dependence on $\gamma_\infty$ in the forward jet. This modest $\gamma_\infty$ dependence likely comes from the fact that the swing is located close to the the true light cylinder, so approximations like the radiation condition start to break down and higher-order corrections of the form of Eq.~\ref{eq:Rsheathpert} are needed.

It is worth noting that resolving $\tilde{R}$ in the counter-jet will be considerably harder in practice than it is for the forward jet. Since the counter-jet is dim, measuring its polarization swing would require higher dynamic range and would also require significant modeling of Faraday and optical depth effects; rays emanating from $\tilde{R}$ in the counter-jet may pass through a disk, increasing the rotation measure and optical thickness. Furthermore, the polarization swing in the counter-jet occurs at much smaller impact parameter than that of the forward jet, requiring higher spatial resolution to be seen.

However, it is promising that the counter-jet polarization swings at roughly the same location independent of the transverse location along the jet. Moreover, these swings depend only modestly on Lorentz factor. While observations of the counter-jet may be more challenging, measurements of $\tilde{R}$ in the counter-jet can therefore be identified with spin and spin alone, thus removing many of the degeneracies highlighted in the previous subsections. Additionally, at high inclinations where we expect the counter-jet swing to be present, Doppler de-boosting becomes less of a concern.


\section{Discussion \& Observational Prospects}
\label{sec:obsforecast}
In this paper, we have investigated the polarization of stationary, axisymmetric black hole jets for arbitrary viewing angle relative to the jet symmetry axis. We have focused solely on optically thin synchrotron emission.

In \S\ref{sec:polfaceon}, we first examined the polarization in the face-on regime, for which the observer views the emission from within the cone of the jet. Drawing on our earlier results \citep{Gelles_2025}, we showed that the polarization of the jet as a function of distance from the black hole depends on how bright the counter-jet is. A bright counter-jet produces a characteristic change in the jet polarization near the light cylinder, which would be a strong probe of black hole spin.   If the counter-jet is faint, the polarization changes more gradually and at larger distances from the black hole (Eq.~\ref{eq:faceontotal}). In both cases, the  polarization change as a function of distance from the black hole provides a pathway to measuring both spin and the jet Lorentz factor with polarimetry.

Then, we examined polarization in the off-axis regime (\S\ref{sec:offaxis}), for which the observer views the emission from outside the jet cone, and for which the forward jet and counter-jet do not overlap in the image. In that case the polarized image breaks down into two sub-components: the spine and the limbs (see Fig. \ref{fig:foldfig}), which are polarized $90^\circ$ with respect to one another far from the black hole. In each sub-component, a spin-dependent polarization swing again emerges (Eqs.~\ref{eq:spineswing} and \ref{eq:Rsheath}). However, the limb swing appears on only one of the two jet edges; when $\gamma_\infty\sin i<1$, the swing appears on the receding edge, and when $\gamma_\infty\sin i>1$, the swing appears on the boosted edge. 

\begin{figure*}[t]
    \centering
    \includegraphics[width=\textwidth]{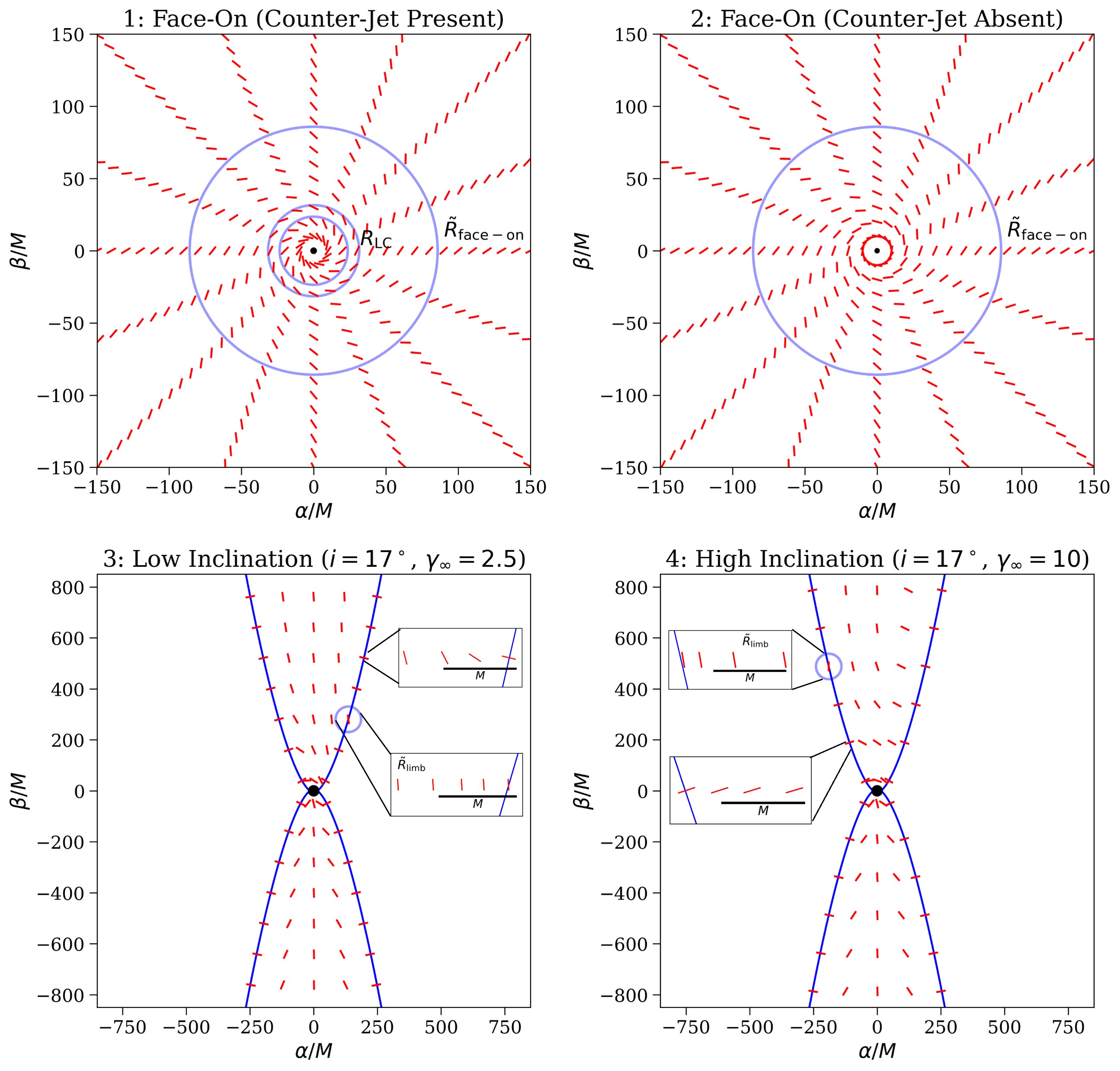}
    \caption{Semianalytic polarization maps in all four regimes. Polarization ticks are plotted in red and track the windup of the EVPA. In each regime, a reliable tracer of spin is plotted in transparent blue: the two lensed images of the light cylinder in Regime 1, $\tilde{R}_{\rm face-on}$ in Regime 2, and $\tilde{R}_{\rm limb}$ in regimes 3 and 4. Crucially, $\tilde{R}_{\rm limb}$ flips from the right side of the image to the left as $\gamma_\infty\sin i$ crosses 1. The counter-jet spine polarization swing (see Eq.~\ref{eq:CJpred}) is not visible in the bottom two panels since the inclination is too low, but we illustrate it separately in Figure~\ref{fig:semipolCJ}.}
    \label{fig:semipolmaps}
\end{figure*}

For observational purposes, we can therefore group polarized jet images into one of four potential categories, each of which has its own qualitatively distinct polarization pattern:
\begin{enumerate}
    \item Face-On Regime (Counter-Jet Present)
    \item Face-On Regime (Counter-Jet Absent)
    \item Low-Inclination Regime $(\gamma_\infty\sin i<1)$
    \item High-Inclination Regime $(\gamma_\infty\sin i>1)$
\end{enumerate}
Composite images of each regime are shown in Figure~\ref{fig:semipolmaps}.

Classifying jet images into one of the four polarimetric regimes requires high dynamic-range polarization data over a range of jet radii. From a polarized image, one can immediately see if the observer falls in the face-on regime based on whether the jet appears axisymmetric or appears to collimate. If the observer is in the face-on regime, one can check for the presence of a counter-jet by searching for the unique polarization swing with distance from the black hole (as in Figure~\ref{fig:faceonCJfig}). If the observer is not in the face-on regime (as is the case for most sources, including many blazars), then one can check whether the observer falls in the low-inclination or high-inclination regime based on the degree of left-right asymmetry in limb polarization (or using independent measurements of $\gamma_\infty$ and $i$). So from polarized images with sufficiently high resolution, classifying the source is feasible. 

\begin{figure*}[t]
    \centering
    \includegraphics[width=0.9\textwidth]{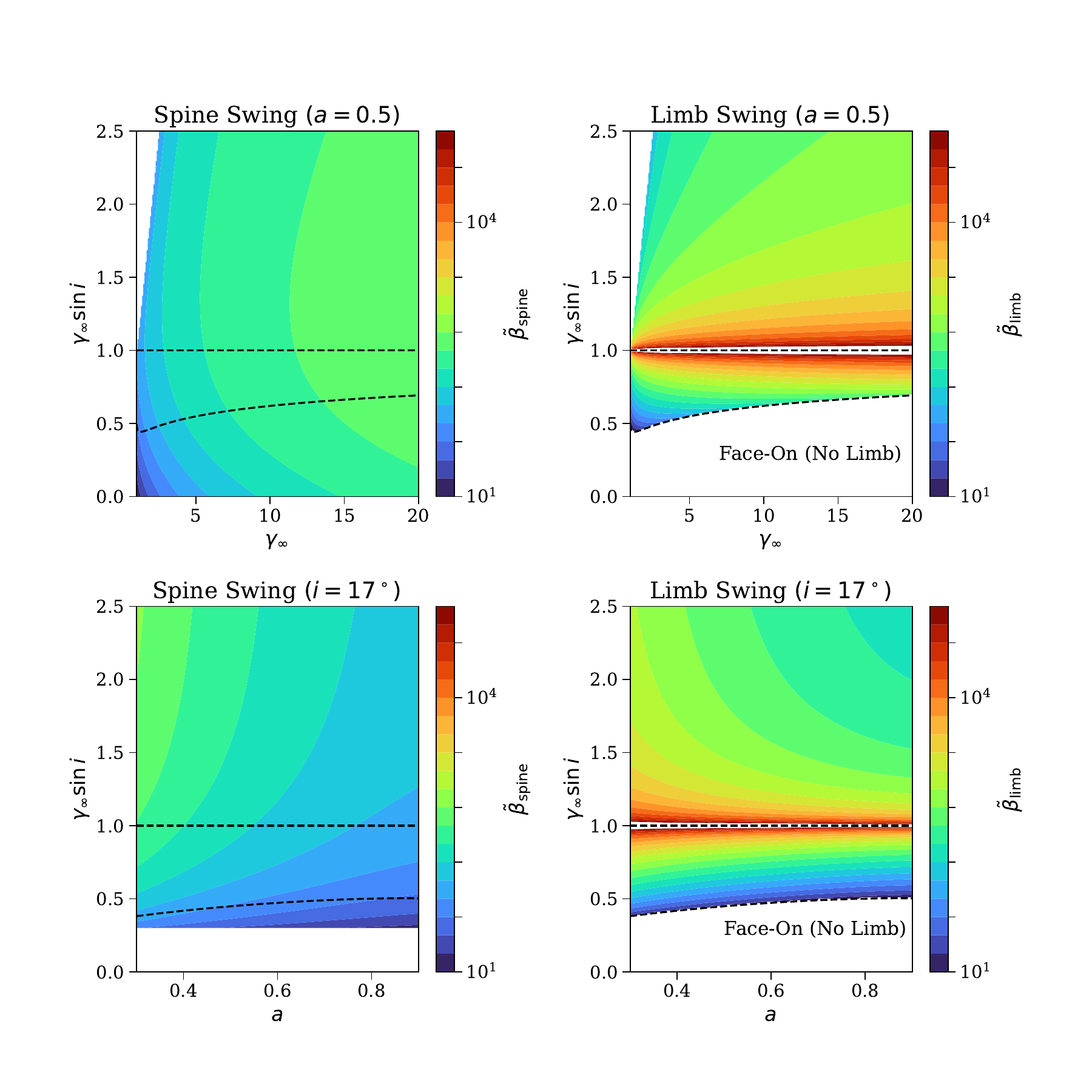}
    \caption{Image coordinate of polarization swing along the spine (left column) and along the relevant limb (right column). For the top row, the black hole spin is fixed, and $\tilde{\beta}$ (in units of $M$) is plotted as a function of $\gamma_\infty$ and $\gamma_\infty\sin i$. For the bottom row, inclination is fixed, and $\tilde{\beta}$ is plotted as a function of black hole spin and $\gamma_\infty\sin i$. Note that $\tilde{\beta}_{\rm limb}$ diverges when $\gamma_\infty\sin i=1$, which is demarcated with the upper black dashed line. The lower black dashed line marks the transition into the face-on regime, where the emission is viewed from within the cone of the jet and the projected limbs disappear from the image.}
    \label{fig:betatildefig}
\end{figure*}

A key result of this paper is our understanding that there is always a characteristic  cylindrical radius $\tilde{R}$ at which the polarization swings $90^\circ$. Since the nature of the polarization swing depends on inclination, we identify $\tilde{R}$ separately for the face-on observer and off-axis observers. In particular, we compute distinct quantities $\tilde{R}_{\rm face-on}$ (Eqs.~\ref{eq:Rtildedef}-\ref{eq:faceonRtilde}), $\tilde{R}_{\rm spine}$ (Eqs.~\ref{eq:spineRtilde}-\ref{eq:spineswing}), and $\tilde{R}_{\rm limb}$ (Eqs.~\ref{eq:Rtildelimb}-\ref{eq:Rsheath}), which are are respectively defined via:\begin{align}
    {\rm EVPA}|_{\tilde{R}_{\rm face-on}}&=\pm 45^\circ\\
    {\rm EVPA}|_{\tilde{R}_{\rm spine}}&=\pm 45^\circ\\
    {\rm EVPA}|_{\tilde{R}_{\rm limb}}&= 0^\circ.
\end{align}
We explicitly show that each of these quantities can then be expressed analytically as a function of spin, inclination angle, and terminal Lorentz factor. $\tilde{R}_{\rm spine}$ grows monotonically with the bulk plasma velocity, as aberration expands the polarimetric signature of the light cylinder. In contrast, $\tilde{R}_{\rm limb}$ is instead maximized for $\gamma_\infty\sin i=1$, at which point $\tilde{R}_{\rm limb}$ moves to infinity. 

In Figure~\ref{fig:betatildefig}, we summarize these dependencies by plotting $\tilde{\beta}$ (the projected image coordinate associated to $\tilde{R}$) along both the spine and limb of a typical black hole jet as a function of both $\gamma_\infty$, $\gamma_\infty\sin i$, and spin. The spine polarization swing (left column) remains continuous all the way into the face-on regime. The limb polarization swing (right column), however, is not well defined when we enter the face-on regime, as the projected edge of the jet disappears. This cutoff is demarcated by the lower black dashed line. Spin dependence --- which is shown in the bottom row --- is strongest when $\gamma\sin i>1$; when $\gamma\sin i\lsim 1$, the swing location is at least as strongly affected by $\gamma_\infty$. For this reason, it is important to use polarimetry to constrain both spin and Lorentz factor \emph{together}.

In all four panels in Figure~\ref{fig:betatildefig}, the polarization swings are visible on scales of $\sim 10-10^4M$, which are precisely the scales of jets that will be probed by millimeter VLBI with the Next Generation Event Horizon Telescope (ngEHT; \citealp{ngEHT1}) and the Black Hole Explorer (BHEX; \citealp{johnson_black_2024}) at high angular resolution in a number of sources. Indeed, in Figure~\ref{fig:incregimes}, we plot several target sources in the parameter space of $\gamma_\infty$ and $\gamma_\infty\sin i$ by using measurements from previous total-intensity studies. We caveat, however, that the error bars on these quantities can be large.

In this plot, sources are separated into the regimes designated above (face-on, low-inclination, high-inclination). Specifically, a point in parameter space $\{\gamma_\infty,\gamma_\infty\sin i\}$ falls into the face-on regime if it produces a polarization swing that the observer views from inside the jet cone: $i<\theta_{\rm swing}$. If instead $i>\theta_{\rm swing}$, then the point falls into either the low-inclination regime or the high-inclination regime simply based on the sign of $\gamma_\infty\sin i-1$. We note that the definition of the face-on regime does depend on the collimation profile and spin of the jet, so we plot it for a range of feasible parameters in Figure~\ref{fig:incregimes}. For this reason, there is a cyan region between the low-inclination and face-on regimes, where the source could be classified into either regime depending on the black hole spin $a$ and collimation index $p$.

We also note that in this plot, all sources satisfy $\gamma_\infty\sin i<2.5$. This is due to the powerful selection effect that relativistic aberration causes jets to be beamed into a cone of half-opening angle $1/\gamma_\infty$. Therefore, jets with $\gamma_\infty\sin i\gg 1$ are completely invisible to us on Earth, though their non-relativistic cores may still be detectable.   In addition, there are a number of radio galaxies which appear to have $\gamma_\infty\sin i$ moderately larger than 1, especially for low values of $\gamma_\infty$ where beaming effects are weaker.

To the right of the plot in Figure~\ref{fig:incregimes}, we include a table of several sources and compute the angular distance of their expected polarization swings. These values are computed for a range of spins $a=0.3-0.9$ and with the fiducial measured values of $\gamma_\infty$ and $i$. This table helps identify which sources are promising targets for jet polarimetry in the near future. For context, the current Event Horizon Telescope has an imaging resolution of ~$20\mu$as \citep{eht_paper1}, while the ngEHT will have $\sim 10\mu$as \citep{ngEHT1} and BHEX will have $\sim 6\mu$as \citep{johnson_black_2024}. Therefore, it is clear that several sources beyond just M87 have spatially resolvable polarimetric swings accessible in the near future, and we expect upgraded interferometric arrays to have the necessary dynamic range to see them.

In principle, measurements of the radius of the polarization swing $\tilde{R}$ can be used to constrain a combination of $a$, $\gamma_\infty$, $i$, and other parameters such as the jet collimation profile $p$. Independent measurements of $i$ and $\gamma_\infty$ or independent measurements of $\tilde{R}$ along the spine and limb can then break degeneracies between these parameters that are present in any single measurement. In practice, we suspect that 
the most promising method for applying our model to observations and assessing the degeneracies in interpreting the data 
will be to fit our model directly to data with a Bayesian inference code or via comparison of our model to the full polarimetric jet data (rather than relying on 1 or 2 summary statistics such as $\tilde R$ measurements).

\begin{figure*}[t]
    \centering
    \includegraphics[width=0.8\textwidth]{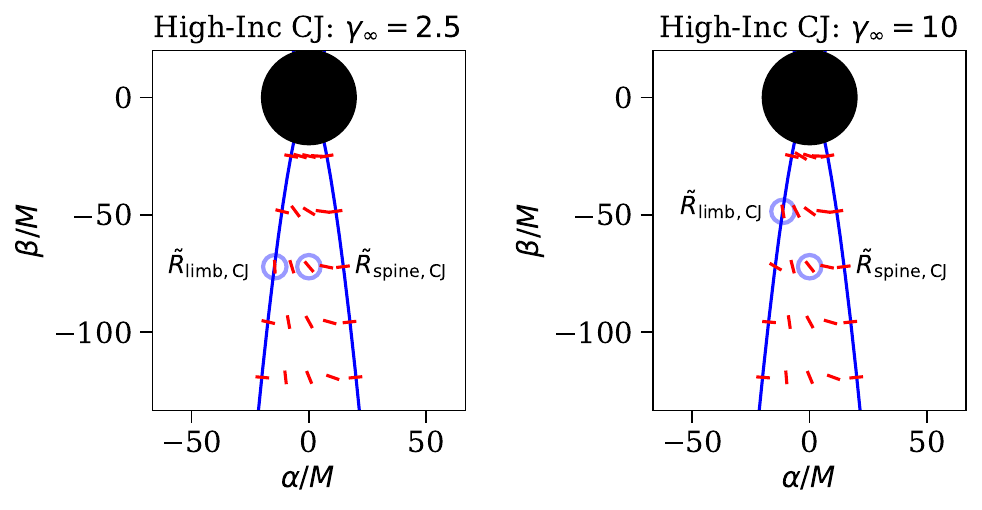}
    \caption{Semianalytic polarization maps of the high inclination $(i=70^\circ;\,\psi=0.15$) counter-jet. For both Lorentz factors shown, a spin-dependent swing emerges along the spine of the counter-jet and the approaching limb of the counter-jet. In both cases, the dependence of this swing on both Lorentz factor and transverse location is significantly weaker than that of the forward jet (see Eq.~\ref{eq:CJpred}).}
    \label{fig:semipolCJ}
\end{figure*}

\begin{figure*}[t]
\centering
\hspace*{-2.8cm}
\begin{overpic}[width=0.85\textwidth]{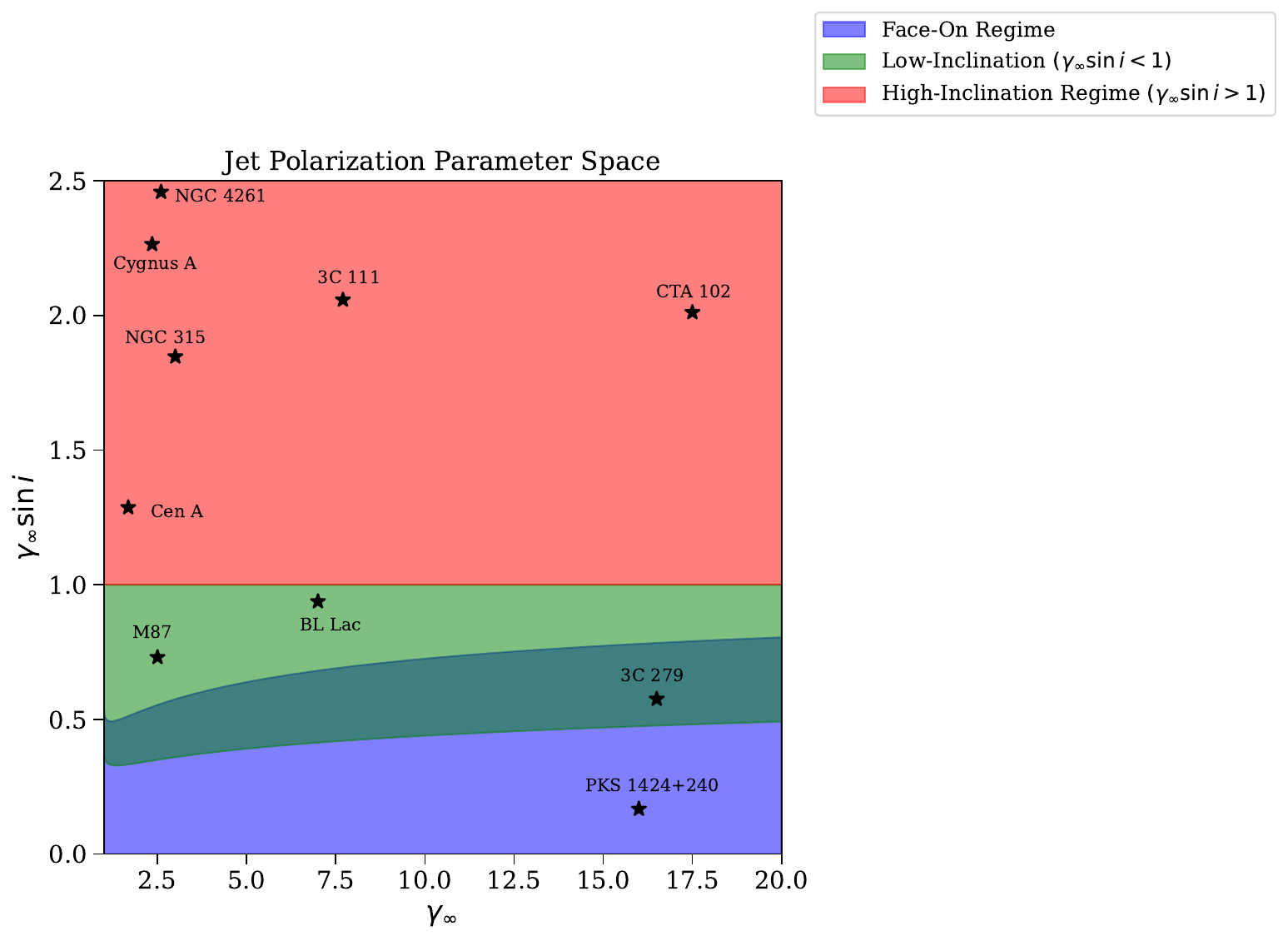}
  \put(63,30){ 
    \begin{minipage}{0.17\textwidth}
    \footnotesize
    \renewcommand{\arraystretch}{1.1}
    \begin{tabular}{lcc p{2.2cm}}
      \textbf{Source} & $\frac{M}{D_A}$ ($\mu$as)  & $\frac{\tilde{\beta}}{D_A}$ ($\mu$as)  & Reference\\ \hline
      M87 & 3.8 & 400-3000 & \cite{mertens_kinematics_2016}
      \\
     NGC 315 & 0.45 & 80-380 & \cite{park_jet_2021,Pilawa_2025}
      \\
     NGC 4261 & 0.5  & 30-150 & \cite{yan2025subparsec}
      \\
      Centaurus A &0.14  & 15-110 & \cite{snios_variability_2019,neumayer2010supermassive}
      \\
      Cygnus A & 0.1 & 7-33 &\cite{Boccardi_2016}
      \\
      BL Lac & 0.001 & 5-30 & \cite{o2009three,titarchuk2017bl}
      \\
      3C 111 & 0.01 & 3-14  & \cite{hovatta_doppler_2009,Beuchert_2018}
      \\
      CTA 102 &0.05&  3-14 & \cite{fromm2015location,li_imaging_2018-1}
        \\
      3C 279 & 0.005 & 1-5.5 & \cite{EHT_3C}
      \\
      PKS 1424+240 & 0.006 & 1-3.5     & \cite{kovalev_looking_2025,Padovani_2022}
    \end{tabular}
    \end{minipage}
  }
\end{overpic}
\caption{On the left, we plot the three inclination regimes used to describe the polarization of jets. The red region corresponds to the \emph{high-inclination regime} with $\gamma_\infty\sin i>1$, the green region corresponds to the \emph{low-inclination regime} with $\gamma_\infty\sin i<1$, and the blue region corresponds to the \emph{face-on regime} with $i<\theta_{\rm swing}$. While the red and green regions are separated by the simple curve $\gamma_\infty\sin i=1$, the separation between green and blue curves will depend more sensitively on model parameters, thus creating the overlap region. Included in this plot are various sources with bright jets and known Lorentz factors/inclinations. On the right, we then list the predicted $\tilde{\beta}$ for each source, assuming a range of spin $a=0.3-0.9$. For all sources, $\tilde{\beta}$ corresponds to the limb swing, except for 3C 279 and PKS 1424+240, for which $\tilde{\beta}$ corresponds to the spine swing. Note that error bars on measured values of $\gamma_\infty$ are often large but are not shown here.}
\label{fig:incregimes}
\end{figure*}

For sources viewed at higher inclination (such as NGC 315 and Cygnus A), constraints on spin can additionally come from polarized observations of the counter-jet (discussed in \S\ref{sec:offaxisCJsec}). At high-inclination, the counter-jet becomes comparable to the forward jet in brightness and develops a spin-dependent polarization swing that is relatively independent of Lorentz factor and transverse location along the jet (Eq.~\ref{eq:CJpred}). Sample images of a high-inclination counter-jet are shown in Figure~\ref{fig:semipolCJ}.

So by combining polarized observations of the forward jet with polarized observations of the counter-jet, we expect to generate even stronger constraints on jet properties and black hole spin.  Indeed, the value of the counter-jet measurements suggests that it would be useful to explicitly search for high inclination targets for VLBI followup by identifying mm-bright cores that lack extended jets (or perhaps only have bright radio lobes far from the host galaxy).

In \S\ref{sec:sheathsec}, we showed that the polarization swing along the limb is typically very localized in radius and is therefore likely to be easily isolated in data, given sufficiently high spatial resolution (e.g., Fig. \ref{fig:sheathcompareinc} \& \ref{fig:sheathcomparegamma}). This bodes well for observational prospects, as many jets in nature are naturally limb-brightened. Indeed, VLBI measurements show that the M87* jet is limb-brightened all the way down to its core \citep{kim2018limb}, a phenomenon which appears to be quite ubiquitous among other AGN too \citep{Takahashi_limb}. Such limb-brightening could be explained, for example, by anisotropy in the electron distribution function controlling the plasma emissivity \citep{Yuh_anisotropy}. With an anisotropic distribution function, none of our results would change, as the observed polarization pattern would still be proportional to $\vec{k}\times\vec{B}$. If anything, the polarization swing along the limb would be \emph{easier} to see, as the polarized intensity there would be enhanced relative to that of the spine.

As we mentioned in \S\ref{sec:truelimb}, the polarization structure is also incredibly rich when analyzed transverse to the jet: near $\tilde{R}_{\rm limb}$, the EVPA develops a kink when plotted as a function of the image coordinate $\varphi$ (see Figure~\ref{fig:evpavsphi}). Ideally, such a kink could be resolved directly in the data. But even with perfect resolution, astrophysical complexities (such as emission from multiple fieldlines) will smear out the kink into something smoother. 

\begin{figure}[h]
    \centering
    \includegraphics[width=0.5\textwidth]{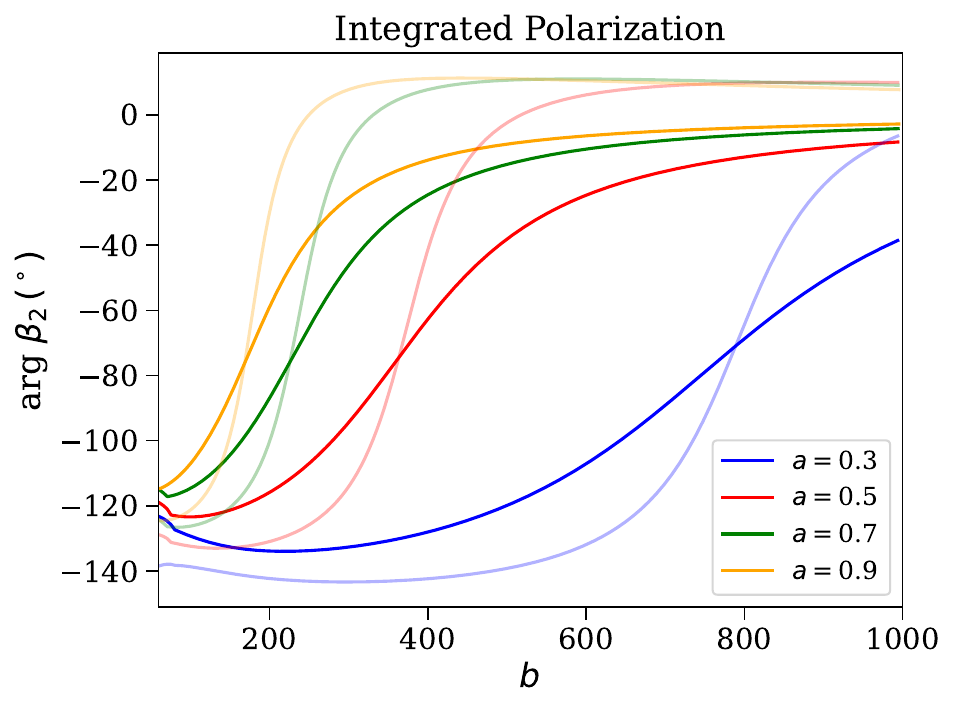}
    \caption{Integrated polarization angle over surfaces of constant impact parameter $(b)$ for an image of a jet with $i=17^\circ$. The result shows a strong spin dependence and captures the effects of aberration as it slows down the polarization evolution with distance from the BH. The solid curves correspond to $\gamma_\infty=2.5$, thus tracking the windup along the receding edge of the jet. The transparent curves correspond to $\gamma_\infty=4.0$, thus tracking the windup along the approaching edge of the jet.}
    \label{fig:argb2fig}
\end{figure}

However, we can still capture the signatures of the transverse polarization structure with the statistic $\arg(\beta_2)$, which measures the EVPA when averaged transverse to the jet axis at a given impact parameter \citep{palumbo_discriminating_2020}: 
\begin{align}
    {\rm arg}(\beta_2)\equiv {\rm arg}\left[\int_{{\rm fixed} \,b} d\varphi\,(Q+iU)e^{-2i\varphi}\right].
\end{align}
This statistic was used to describe the EVPA of face-on images in \cite{Gelles_2025}. It can be used to describe the azimuth-averaged EVPA in off-axis images as well, despite the fact that the polarization pattern is no longer axisymmetric. We plot $\arg(\beta_2)$ for two different Lorentz factors in a sample jet configuration Figure~\ref{fig:argb2fig}.
We see that this statistic is powerful because it produces a spin-dependent swing that accounts for the transverse structure of the jet. We therefore expect $\arg(\beta_2)$ to play a significant role in measuring spin from polarized images of black hole jets.

The focus of this paper has intentionally been on idealized, time-steady, axisymmetric jet models. In the context of such models, we have identified a number of ways that jet polarization observations at radii of $\sim 10-10^4 GM/c^2$ can provide unique and powerful constraints on the rotation of the jet (receding vs approaching), our viewing angle $i$ in a given system, jet Lorentz factor $\gamma_\infty$, and (perhaps most excitingly) black hole spin. These promising observational signatures motivate additional followup to quantify the robustness of our proposed diagnostics in the context of more realistic jets. One caveat, for example, is that
measurements of polarization along the spine and along the limbs may actually be probing different fieldlines, rather than the single field line that we have focused on in this paper. Indeed, many blazars show apparent velocity stratification between spine and sheath, with a fast inner layer and a slow outer layer \citep{Ghisellini_2005,Pushkarev_2005_spine,Ito_2014,Sikora_2016}. This indicates that our fiducial choice of a single field line with $\psi=1$ (which would have the same kinematics along the spine and limb) may need to be modified to capture the complete transverse structure observed in real systems.  

More broadly, one lesson of our models in this paper is that the superposition of counter-jet and forward jet or foreground and background emission can be very important in shaping the net polarization observed (e.g., Figues \ref{fig:cjcomparefig},  \ref{fig:spinecompareinc}, \& \ref{fig:sheathcompareinc}). In reality, the net polarization in a given region of the jet will also depend on a superposition of emission from different fieldlines. And which field lines have the highest emissivity is likely to vary spatially and temporally, due to local magnetic dissipation and particle acceleration physics. It remains to be determined how this affects the jet polarimetry signatures identified here.

Faraday rotation will also be important to consider in future modeling efforts. While our definition of $\tilde{R}$ is anchored to an absolute value of the EVPA, it can be modified so that $\tilde{R}$ corresponds to ``halfway along the EVPA swing," which is a definition that remains invariant under external Faraday rotation. Helical magnetic fields can also produce strong rotation measure gradients across jets that can significantly rotate the EVPA of the spine relative to that of the limbs \citep{asada2002helical,zavala2005faraday,broderick2010parsec,gabuzda2015transverse}. While this is less likely to be important at higher frequencies  and multifrequency observations can correct for the effects of Faraday rotation, single frequency polarimetry might still be sufficient; as long as the swings along the spine and limbs are measured separately and the rotation measure does not depend strongly on $Z$, then the effects of Faraday rotation will largely decouple from our proposed polarimetric signatures of spin and Lorentz factor.

A useful theoretical exercise will be to compare our model predictions to ray-traced images of jets produced in 3D GRMHD simulations. These will probe the extent to which a clear measurement of the polarization swing radius $\tilde{R}$ can be extracted from non-axisymmetric, time-dependent mock data (even though such mock data will likely not capture the full complexity of real jets).   
It will also be valuable to generate synthetic data and perform realistic image reconstructions or direct fits of our model to data. The M87* polarimetric data is already extremely rich, and recent measurements have shown polarized images of the jet with the kind of limb asymmetry we have predicted (albeit at surprisingly large scales; \citealp{park2025helicalmagneticfieldaccelerationcollimation}).  

Another important theoretical step will be generalizing our work to the optically thick regime. In this paper, we have neglected radiative transfer effects, as appropriate for observing frequencies where synchrotron self-absorption is negligible ($\gtrsim 86$ GHz in M87; \citealt{lee_interferometric_2016}). But in the optically thick regime, the radiative transfer will become crucial; polarization will be preferentially absorbed perpendicular to the magnetic field and hence leave behind an EVPA that is $90^\circ$ rotated from what we predicted here \citep{Gabuzda_radiation}. Optical depth will likely be particularly important for higher spin and more distant sources where the polarization changes take place closer to the core.


\begin{acknowledgements}
    We thank Dominic Chang, Charles Gammie, Michael Johnson, Yuri Kovalev, Alexandru Lupsasca, Ramesh Narayan, Daniel Palumbo, Dom Pesce, Aditya Tamar, Paul Tiede, and George Wong for helpful discussions. ZG was supported by a National Science Foundation Graduate Research Fellowship and benefited from Harvard computing resources. AC was supported by the Princeton Gravity Initiative. This work was supported in part by a Simons Investigator Award to EQ and benefited from EQ's stay at the Aspen Center for Physics, which is supported by National Science Foundation grant PHY-2210452.
\end{acknowledgements}

\appendix
\section{Glossary}
\label{app:glossary}
\begin{description}
  \item[Forward Jet] Emission from the part of the jet  beamed toward the observer.
  \item[Counter-jet] Emission from the part of the jet lobe beamed away from the observer.
  \item[Foreground] Emission from the region between the observer and the jet axis $(Y<0)$. See Figure~\ref{fig:inclinationfig}.
  \item[Background] Emission from the region behind the jet axis $(Y>0)$. See Figure~\ref{fig:inclinationfig}.
  \item[Spine] Emission near the vertical axis of the projected jet (i.e. $X\approx 0$). See Figure~\ref{fig:foldfig}.
  \item [Limb] Emission near the edges of the projected jet (i.e. $Y\approx 0$).  See Figure~\ref{fig:foldfig}. 
  \item [True Limb] The \emph{exact} projected edge of the jet, defined as the location where $\vec{k}\cdot\nabla \psi=0$. See Figure~\ref{fig:foldfig}. 
  \item [$R_{\rm LC}$] The cylindrical radius of the light cylinder --- the point at which the magnetic fieldlines rotate at the speed of light and the magnetic field becomes toroidally dominated.
   \item [$\tilde{R}_{\rm face-on}$] The cylindrical radius where the EVPA of a face-on jet swings through $\pm 45^\circ$. Spin dependence encoded in Eq.~\ref{eq:faceontotal}.
  \item [$\tilde{R}_{\rm spine}$] The cylindrical radius where the spine EVPA swings through $\pm 45^\circ$. Spin dependence encoded in Eq.~\ref{eq:spineswing}.
  \item [$\tilde{R}_{\rm limb}$] The cylindrical radius where the limb EVPA swings through $0^\circ$. Spin dependence encoded in Eq.~\ref{eq:Rsheath}.
\end{description}
\section{Coordinates}
\label{app:coordinates}
Throughout this work, we use two different sets of orthonormal basis vectors: Cartesian basis vectors $\{e_X,e_Y,e_Z\}$ and cylindrical basis vectors $\{e_R,e_{\hat{\phi}},e_Z\}$, which are related via\begin{align}
   e_R&=\cos\phi\, e_X+\sin \phi\, e_Y,\quad e_{\hat{\phi}}=-\sin\phi \,e_X+\cos\phi\, e_Y,\quad e_Z=e_Z.
\end{align}
The ``hat" on the basis vector $e_{\hat{\phi}}$ serves to emphasize that this vector is already unit normalized. Therefore, vectors in both coordinate systems are raised and lowered using the standard Euclidean metric $g_{ij}=\delta_{ij}$. This convention is most convenient in flat space, where the full Lorentzian metric can easily be written in an orthonormal basis at any given point.

This stands in contrast with the unit vectors employed in \cite{Gelles_2025}, which instead used the basis vector $e_\phi\equiv Re_{\hat{\phi}}$. This alternative approach is necessary when performing computations in curved space, where the metric is naturally expressed in an unnormalized basis.

\section{The Face-On Regime}
\label{app:faceon}
Here, we derive Eq.~\ref{eq:faceontotal} --- the location of the polarization swing for the face-on observer. To do so, we begin with Eq.~\ref{eq:evpafaceon}. To simplify this formula, we will expand the stream function used in \cite{Gelles_2025} in large $R$, and here we will not plug in $\psi=1$ so as to allow for maximum generality. This gives the following useful approximations:\begin{align}
    \frac{B^R}{B^{\hat{\phi}}}&= \frac{2-p}{2}\frac{R_{\rm LC}}{Z}+...= \frac{2-p}{2}R_{\rm LC}\left(\frac{R}{\sqrt{2\psi}}\right)^{-\frac{2}{2-p}}+...
    \\
    \label{eq:BphiEReq}
   \frac{-B^{\hat{\phi}}}{E_R}&=1+p\psi Z^{-p}+...=1+p\psi\left(\frac{R}{\sqrt{2\psi}}\right)^{-\frac{2p}{2-p}}+...
\end{align}
Plugging into Eq.~\ref{eq:faceontotal} and expanding the denominator, this gives\begin{align}
    \tan({\rm EVPA})|_{\rm face-on}&\approx -\frac{2-p}{2}R_{\rm LC}\left(\frac{R}{\sqrt{2\psi}}\right)^{-\frac{2}{2-p}}\left[\frac{1}{1-\frac{v}{v_{\rm FF}}+p\psi \left(\frac{R}{\sqrt{2\psi}}\right)^{-\frac{2p}{2-p}}}\right].
\end{align}
In the non-relativistic, limit, $1-v/v_{\rm FF}$ remains finite, and the sub-leading $R$ term in the denominator can be ignored. In the ultra-relativistic limit, $v/v_{\rm FF}\approx 1$, and the sub-leading $R$ term in the denominator is crucial. Inverting for $R$ in each of these two regimes gives\begin{align}
    \tan({\rm EVPA})|_{\tilde{R}_{\rm face-on}}=0\Longrightarrow \tilde{R}_{\rm face-on}&=\begin{cases}
        \sqrt{2\psi}\left[\frac{1-p/2}{1-v_\infty}R_{\rm LC}\right]^{1-p/2},&\gamma_\infty\sim 1
        \\
        \sqrt{2\psi}\left[\frac{2-p}{2p\psi}R_{\rm LC}\right]^{\frac{1-p/2}{1-p}},&\gamma_\infty \gg 1,
    \end{cases}
\end{align}
which is Eq.~\ref{eq:faceontotal}.

\section{Polarization Identities}
\subsection{Frame-Independent Expression}
\label{app:polidentities}
Here, we derive Eq.~\ref{eq:polcross}. To do so, we recall that $\vec{f}\propto \vec{k}\times\vec{B}$ in the fluid frame. This can be expressed covariantly as \citep{Hou_2024,Gelles_2025}\begin{align}
    f^\mu\propto \epsilon^{\mu\nu\rho\sigma} u_\nu k_\rho B_\sigma,
\end{align}
with $u_\rho=\gamma(-1,v_i)$ the four-velocity. Now, given the above formula, we only need to worry about the component of the four-velocity perpendicular to the magnetic field. In MHD, one can always write\begin{align}
    v^\mu_\perp\propto v^\mu_{\perp,\rm FF}=-\frac{\epsilon^{\mu\nu\rho\sigma}\eta_\nu E_\rho B_\sigma}{B^2},
\end{align}
and so\begin{align}
    u^\mu_\perp\propto \eta^\mu-\left(\frac{v^\perp}{v_{\rm FF}^\perp}\right)\frac{\epsilon^{\mu\nu\rho\sigma}\eta_\nu E_\rho B_\sigma}{B^2},
\end{align}
where $\eta^\mu=(1,0,0,0)$ is the normal vector. Therefore,\begin{align}
    f^\mu&\propto\epsilon^{\mu\nu\rho\sigma}\left[\eta_\nu-\left(\frac{v^\perp}{v_{\rm FF}^\perp}\right)\frac{\epsilon_{\nu\alpha\beta\gamma}\eta^\alpha E^\beta B^\gamma}{B^2}\right]k_\nu B_\sigma.
\end{align}
Evaluating the spatial components only in flat space with $k_t=-1$, we get\begin{align}
    \vec{f}\propto \vec{k}\times\vec{B}-\left(\frac{v^\perp}{v_{\rm FF}^\perp}\right)\frac{(\vec{E}\times\vec{B})\times\vec{B}}{\vec{B}^2}=\vec{k}\times\vec{B}-\left(\frac{v^\perp}{v_{\rm FF}^\perp}\right)\left[\vec{E}-\frac{\vec{E}\cdot\vec{B}}{\vec{B}^2}\right]=\vec{k}\times\vec{B}-\left(\frac{v^\perp}{v_{\rm FF}^\perp}\right)\vec{E}.
\end{align}
This gives
\begin{align}
\label{eq:step1}
    \vec{f}\propto \vec{k}\times\vec{B}-\vec{v}\times\vec{B}.
\end{align}
Now, in \cite{Gelles_2025}, we showed that the four-velocity in cold, axisymmetric (GR)MHD is very well approximated by the force-free drift velocity with a smooth cap on the Lorentz factor:\begin{align}
    \vec{v}_{\rm MHD}^\perp\approx \left(\frac{v^\perp}{v_{\rm FF}^\perp}\right)\vec{v}_{\rm FF}^\perp,\quad \gamma_{\rm MHD}=(\gamma_\infty^{-2}+\gamma_{\rm FF}^{-2})^{-1/2},\quad \vec{v}_{\rm FF}^\perp&=\frac{\vec{E}\times\vec{B}}{\vec{B}^2}.
\end{align}
Therefore,\begin{align}
\label{eq:driftidentity}
    \vec{v}_{\rm MHD}\times\vec{B}&=\left(\frac{v^\perp}{v_{\rm FF}^\perp}\right)\frac{(\vec{E}\times\vec{B})\times\vec{B}}{\vec{B}^2}=-\left(\frac{v^\perp}{v_{\rm FF}^\perp}\right)\left[\vec{E}-\frac{\vec{E}\cdot\vec{B}}{\vec{B}^2}\right]=-\left(\frac{v^\perp}{v_{\rm FF}^\perp}\right)\vec{E},
\end{align}
where we applied both the BAC-CAB rule and the degeneracy requirement $\vec{E}\cdot\vec{B}=0$. Plugging this identity into Eq.~\ref{eq:step1} gives\begin{align}
    \vec{f}\propto \vec{k}\times\vec{B}+\left(\frac{v^\perp}{v_{\rm FF}^\perp}\right)\vec{E}.
\end{align}
In the lab frame, the electric field is given by\begin{align}
    \vec{E}|_{\rm lab}&=R\vec{\Omega}_F\times\vec{B},\quad \vec{\Omega}_F=\Omega_F\hat{\phi}.
\end{align}

\subsection{Limb Swing Location}
\label{app:pertsheath}
Here, we briefly explain how we derive the first perturbative corrections to the location of the limb polarization swing $\tilde{R}_{\rm limb}$ (Eq.~\ref{eq:Rsheath}). 

We start with Eq.~\ref{eq:fRsheathbefore} and substitute in the first correction to the radiation condition from Eq.~\ref{eq:BphiEReq}. Next, we need to include corrections that reflect the full evolution of the velocity profile. Indeed, throughout this work, we employ the smooth velocity prescription from \cite{Gelles_2025}:\begin{align}
    \gamma_{\rm MHD}\approx (\gamma_{\rm FF}^{-2}+\gamma_{\infty}^{-2})^{-1/2}=\frac{\sqrt{v_\infty^2+v_{\rm FF}^2-1}}{v_{\rm FF}}.
\end{align}
We can then expand\begin{align}
    v_{\rm FF}&=\frac{\vec{E}\times\vec{B}}{\vec{B}^2}\approx \frac{E^{R}}{B^{\hat{\phi}}}\approx 1-p\psi\left(\frac{R}{\sqrt{2\psi}}\right)^{-\frac{2p}{2-p}},
\end{align}
thus giving the leading order spatial dependence to the Lorentz factor. 

The final step is to expand around $\tilde{R}_0$, writing\footnote{Note that another term in the expansion formally appears when $p<2/3$, but we find that Eq.~\ref{eq:Rsheathpert} is still accurate in that case.}\begin{align}
    R&=\tilde{R}_0+\mathcal{A}\tilde{R}_0^{\frac{2-3p}{2-p}}+\mathcal{B}\tilde{R}_0^{-1}+\mathcal{O}(\tilde{R}_0^{-\frac{4p}{2-p}})
\end{align}
Evaluating $f^R$ under all of these substitutions and employing the binomial theorem to expand expressions with non-integer exponents:\begin{align}
    (1+x)^p=\suml_{k=0}^\infty\frac{\Gamma(p+1)}{\Gamma(k+1)\Gamma(p-k+1)}x^k,
\end{align}
we obtain\begin{align}
    f^R&=\left[\mathcal{A}-\frac{2^{-\frac{2(1-p)}{2-p}}(3-2p)\psi^{\frac{2}{2-p}}}{v_\infty(v_\infty-\cos i)}\right]\tilde{R}_0^{\frac{-2p}{2-p}}+\left[\mathcal{B}-\frac{1}{2\Omega_F^2\gamma_\infty^{2}v_\infty(v_\infty-\cos i)}\right]\tilde{R}_0^{-2}+\mathcal{O}(\tilde{R}_0^{-\frac{-2-3p}{2-p}}).
\end{align}
By definition, we have $f^R|_{\tilde{R}}=0$, meaning that we must take
\begin{align}
    \mathcal{A}&=\frac{2^{-\frac{2(1-p)}{2-p}}(3-2p)\psi^{\frac{2}{2-p}}}{v_\infty(v_\infty-\cos i)}
    \\
    \mathcal{B}&=\frac{1}{2\Omega_F^2\gamma_\infty^{2}v_\infty(v_\infty-\cos i)}.
\end{align}

\subsection{True Limb}
\label{app:truelimb}
Here, we derive the fact that $f=0$ when $R=\tilde{R}$ on the true limb. To do so, we first observe that along the true limb, $f^\phi|_{\rm TL}=0$:\begin{align}
    f^\phi|_{\rm TL}&=(\vec{k}\times\vec{B})^\phi+\left(\frac{v^\perp}{v^\perp_{\rm FF}}\right)E^\phi=(\vec{k}\times\vec{B})^\phi=B^Rk^Z-B^Zk^R=-R(k^Z\p_Z\psi+k^Z\p_Z\psi)\propto \vec{k}\cdot\nabla\psi=0,
\end{align}
where the last step used the definition of the true limb (Eq.~\ref{eq:TLcondition}). This in turn implies that $f^t|_{\rm TL}=0$ as well:\begin{align}
    f^{t}|_{\rm TL}&\propto \epsilon^{t\mu\nu\rho}u_\mu^\perp k_\nu B^\rho\propto -u_\phi^\perp[k_\theta B^r-k_rB^\theta]-k_\phi[u_r^\perp B^\theta-u_\theta^\perp B^r]-B^\phi[k_ru_\theta^\perp-k_\theta u_r^\perp]
    \propto (\vec{v}^\perp\times\vec{k})_\phi\\&\propto[(\vec{E}\times\vec{B})\times\vec{k}]_\phi=E_\phi(\vec{B}\cdot\vec{k})-B^\phi(\vec{k}\cdot\vec{E})=B^\phi(\vec{k}\cdot\vec{E})\propto k^\theta B^r-k^r B^\theta=0.
\end{align}
In fact, along the true limb, not only are the $t$ and $\phi$ components of $f$ zero in the lab frame, but they are also zero in the fluid frame. The former holds by definition, and the latter holds because\begin{align}
    f_{(\phi)}|_{\rm TL}&=e^\mu_{(\phi)} f_\mu|_{\rm TL}=e^{t}_{(\phi)}f_t|_{\rm TL}+e^{\phi}_{(\phi)} f_\phi|_{\rm TL}=0,
\end{align}
as $e^r_{(\phi)}=e^\theta_{(\phi)}=0$ (here, $e^\mu_{(a)}$ is the tetrad used to transform tensors from the lab frame to the fluid frame; see \citealp{dexter_public_2016} for more detail). So along the true limb, the polarization is purely poloidal \emph{in all frames}.

Now let us evaluate Eq.~\ref{eq:polsheath} along the TL. Since the polarization is purely poloidal in all frames, then $f^R|_{\rm TL}\propto f^{(R)}|_{\rm TL}$ and $f^{(Z)}|_{\rm TL}\propto f^Z|_{\rm TL}$. By definition, the numerator of Eq.~\ref{eq:polsheath} vanishes at $\tilde{R}$, so \begin{align}
    0&=f^R|_{\tilde{R},\rm TL}=f^{(R)}|_{\tilde{R},\rm TL}=[k^{(\phi)}B^{(Z)}-k^{(Z)}B^{(\phi)}].
\end{align}
This can be plugged into the expression for $f^Z$ to find:\begin{align}
    f^Z|_{\tilde{R},\rm TL}\propto f^{(Z)}|_{\rm swing,TL}=[k^{(R)}B^{(\phi)}-k^{(\phi)}B^{(R)}]=\left[k^{(R)}B^{(\phi)}-\frac{k^{(Z)}B^{(R)}B^{(\phi)}}{B^{(Z)}}\right]=0,
\end{align}
where in the final step we substituted\begin{align}
    k^{(R)}=\frac{k^{(Z)}B^{(R)}}{B^{(Z)}}
\end{align}
from the condition $f^{(\phi)}|_{\rm TL}=0$. Thus, $f$ completely vanishes at $\tilde{R}$ along the TL.

\software{
\texttt{kgeo} \citep[][]{chael_kgeo_2023},
eht-imaging library \citep{chael_eht-imaging_2018}
, 
Numpy \citep{harris_array_2020},
Matplotlib \citep{hunter_matplotlib_2007}}
\bibliographystyle{aasjournal} 


\end{document}